\documentclass[paper]{JHEP3}
\JHEPspecialurl{http://jhep.sissa.it/JOURNAL/JHEP3.tar.gz}
\usepackage{epsfig,multicol}
\usepackage{cite}       
\usepackage{amsmath,amssymb}  
\usepackage{bm}   
\usepackage{enumerate} 
\usepackage{wrapfig,float} 
\usepackage{longtable} 
\newcommand {\beq}{\begin{equation}}
\newcommand {\eeq}{\end{equation}} 
\newcommand {\beqa}{\begin{eqnarray}}
\newcommand {\eeqa}{\end{eqnarray}}

\newcommand {\tr}{{\rm tr\,}}

\newcommand {\ee}{\mbox{e}}

\newcommand{\tX}{\tilde{X}}

\newcommand{\tpsi}{\tilde{\psi}}

\newcommand{\tA}{\tilde{A}}

\newcommand{\tc}{\tilde{c}}
\newcommand{\bc}{\bar{c}}
\newcommand{\tbc}{\tilde{\bar{c}}}

\newcommand{\al}{\alpha}
\newcommand{\be}{\beta}
\newcommand{\gm}{\gamma}
\newcommand{\dl}{\delta}

\newcommand{\ps}{\psi}

\newcommand{\oot}{\frac{1}{2}}

\newcommand{\oof}{\frac{1}{4}}
\newcommand{\ooN}{\frac{1}{N}}

\newcommand{\B}{\Big}

\newcommand{\intdt}{\int_0^\be \!dt\,}

\preprint{KEK-TH-1193}
\title{
High temperature expansion in supersymmetric 
matrix quantum mechanics 
}
 
\author{ 
Naoyuki Kawahara${}^{ab}$\,,
Jun Nishimura${}^{ac}$ 
and Shingo Takeuchi${}^{c}$ 
\vspace*{0.5cm} \\  
\llap{$^a$}High Energy Accelerator Research Organization (KEK),\\
Tsukuba, Ibaraki, 305-0801, Japan  \\
\llap{$^b$}Department of Physics, Kyushu University,
Fukuoka 812-8581, Japan \\ 
\llap{$^c$}Department of Particle and Nuclear Physics,\\
Graduate University for Advanced Studies (SOKENDAI),\\
Tsukuba, Ibaraki, 305-0801, Japan
\vspace*{0.5cm} \\ 
\email{kawahara@post.kek.jp,
jnishi@post.kek.jp, shingo@post.kek.jp}
}

\abstract{
We formulate the high temperature expansion 
in supersymmetric matrix quantum mechanics
with 4, 8 and 16 supercharges.
The models can be obtained by dimensionally reducing
${\cal N}=1$ U($N$) super Yang-Mills theory in $D=4,6,10$ 
to 1 dimension, respectively.
While the non-zero frequency modes become weakly coupled
at high temperature,
the zero modes remain strongly coupled.
We find, however, that  
the integration over the zero modes
that remains after integrating out all the
non-zero modes perturbatively, 
reduces to the evaluation of 
connected Green's functions
in the bosonic IKKT model.
We perform Monte Carlo simulation
to compute these Green's functions,
which are then used to obtain the
coefficients of the high temperature expansion
for various quantities up to the next-leading order.
Our results nicely reproduce the asymptotic behaviors
of the recent simulation results at finite temperature.
In particular, the fermionic matrices,
which decouple at the leading order,
give rise to substantial 
effects
at the next-leading order, 
reflecting finite temperature behaviors 
qualitatively different from the 
corresponding models without fermions.
}


\begin{document} 

%==================================================================
\section{Introduction}
\label{sec:Intro} 
%==================================================================

%%%

Recently large-$N$ gauge theories are playing 
increasingly
%more and more  
important roles in string theory.
% from two different points of view.
%One is that they  
One of the crucial discoveries was that
U($N$) gauge theory appears
as a low energy effective theory
\cite{Witten:1995im} of a stack of $N$ D-branes 
\cite{Polchinski:1995mt}
in string theory.
This led to various interesting 
conjectures.
% concerning large $N$ gauge theories.
For instance,
% it is conjectured that
large-$N$ gauge theories obtained
by dimensionally reducing 10d U($N$) super Yang-Mills
theory to $0,1,2$ dimensions
are conjectured to provide non-perturbative formulations of 
superstring/M theories \cite{BFSS,IKKT,DVV}.
%These are called the IIB matrix model \cite{IKKT},
%Matrix theory \cite{BFSS},
%and matrix string theory \cite{DVV}, respectively.

Another type of conjectures
%, which is studied extensively during the decade,
asserts the duality between strongly coupled 
large-$N$ gauge theory and
%have dual descriptions in terms of 
weakly coupled supergravity.
In the AdS/CFT correspondence
\cite{Maldacena:1997re},
%,Witten:1998qj,Gubser:1998bc},
for instance,
it is conjectured that
4d $\mathcal{N}=4$ ${\rm U}(N)$ 
super Yang-Mills theory is dual to 
the type IIB supergravity on ${\rm AdS}_5 \times {\rm S}^5$.
This duality is generalized to
the finite temperature setup
%, it was argued that
%the deconfinement phase transition in the gauge theory
%corresponds to the Hawking-Page transition in the gravity theory
\cite{Witten:1998zw}
and 
%This has been further extended 
to non-conformal gauge theories
\cite{Itzhaki:1998dd}.
Motivated by such dualities,
large-$N$ gauge theories at finite temperature 
\cite{Aharony,Aharony:2004ig,Aharony4,%
wadia}
have been studied intensively.
%, which revealed
%interesting connections to the black-hole physics.
%
%an active field of research.

%although it becomes very time-consuming in the presence of 
%fermions \cite{Krauth:1998xh,Krauth:1998yu,%
%Ambjorn:2000bf,Ambjorn:2000dx,sign}.
%
%In this circumstances, it is
%
%As a first step towards the complete understanding 
%of the phase diagram, it is 
%very useful to consider the high temperature limit.

Monte Carlo simulation is expected to be a powerful approach
to explore the phase diagram of
large-$N$ gauge theories.
%from first principles.
Indeed there was a remarkable progress in this direction
recently. Supersymmetric matrix quantum mechanics
have been 
studied by Monte Carlo simulation for the first time 
\cite{Hanada:2007ti,Catterall:2007fp,Anagnostopoulos:2007fw}.
Ref.\ \cite{Anagnostopoulos:2007fw}, in particular,
deals with the model with 16 supercharges,
which may be viewed\footnote{The model is formally the same
as the non-perturbative
formulation \cite{BFSS} of the M theory
% \cite{Witten:1995ex}
although the large-$N$ limit should be taken in a different way.
}
%Note, in particular, that the high temperature behavior
%we study here 
%does not correspond to the high behavior of the M theory.
as the low energy effective theory of $N$
D0-branes in the type IIA superstring theory \cite{Witten:1995im}.
The Monte Carlo results confirmed
the gauge/gravity duality
from first principles.
%
%provided the microscopic origin of
Unlike in the bosonic model
\cite{latticeBFSS,Aharony:2004ig,Kawahara:2007fn},
no phase transition was observed
at finite temperature,
which is consistent with the prediction based
on the gauge/gravity duality 
\cite{Barbon:1998cr,Aharony4}.
Moreover, the internal energy at low temperature agreed
with that of the non-extremal black hole described
by the dual geometry.
%%on the gravity side 
%in terms of 
%has been reproduced from the strongly coupled gauge theory.
This implies in particular that
%Moreover, 
the Bekenstein-Hawking entropy of the black hole
is given a microscopic origin
in terms of
the open strings attached to the constituent $N$ D0-branes.
Unlike in Strominger-Vafa's result \cite{Strominger:1996sh}
for the extremal black hole, which relied on 
the supersymmetric non-renormalization theorem,
the agreement has been found
by studying the strongly coupled dynamics of the D0-brane effective
theory directly.
%% This can be compared
%% with Strominger-Vafa's calculation \cite{Strominger:1996sh}
%% for the extremal black hole,
%% where the supersymmetric non-renormalization theorem
%% was used to reduce the problem to the counting of 
%% microscopic states in the free theory.
%% In the present case of a non-extremal black hole,
%% the supersymmetric non-renormalization theorem
%% is not valid any more.
%% However, we were able to study
%% the strongly coupled dynamics of the D0-brane effective
%% theory directly, and 
%% obtained agreement with the black-hole thermodynamics.
See refs.\ \cite{KLL,Itzhaki:1998dd}
for earlier works,
which connect the supersymmetric matrix quantum mechanics
at finite temperature
to the black-hole physics through the 
gauge/gravity duality.
%This conclusion is obtained from the behavior of 
%Moreover, the free energy obtained by the Gaussian 
%expansion method shows a power-law dependence on the
%temperature at low

In this paper we formulate the high temperature expansion
in the supersymmetric matrix quantum mechanics.
While the low temperature behavior of the theory
describes the classical black hole,
the high temperature behavior is expected to
describe hot strings \cite{Horowitz:1996nw}.
%Generalizing the most interesting case with
%16 supercharges,
%maximum supersymmetry, 
%Here we consider
We study the models with $4,8,16$ supercharges
that can be obtained by dimensionally reducing
${\cal N}=1$ U($N$) super Yang-Mills theory in $D=4,6,10$ 
to 1 dimension.
The high temperature limit of the 
$D=10$ case \cite{BFSS}
%Matrix Theory 
for $N=2$ has been studied 
in ref.\ \cite{Bal-Sathiapalan}.
As observed there and also 
in refs.\ \cite{Aharony:2004ig,Kawahara:2007nw},
only the bosonic zero modes survive
at the leading order,
%all the non-zero frequency modes are decoupled, and 
%we are left with 
%which 
and their dynamics are governed by the bosonic part of
the IKKT \cite{IKKT} matrix model.
In order to see the effects of the fermions,
we proceed to the next-leading order.
After integrating out the weakly-coupled 
non-zero frequency modes perturbatively, 
we find that the 
remaining integration over the zero modes
reduces to the evaluation of 
connected Green's functions
in the bosonic IKKT model.
This can be done by Monte Carlo simulation
with much less effort than simulating
the supersymmetric models at finite temperature directly.
In particular, we are able to make a reliable 
large-$N$ extrapolation using the data for
$N$ up to 32.
%see clear large $N$ scaling behavior and .
% for matrix size $16 \le N \le 32$
%
%First we integrate out the weakly-coupled 
%non-zero modes perturbatively.
%Then the 
%calculation of various quantities
%
%such as the Polyakov line, the internal energy
%and the extent of the eigenvalue distribution
%
%% then reduces to the evaluation of 
%% various connected Green's functions 
%% in the bosonic IKKT model, which can be done
%% by Monte Carlo simulation. 
%
%reduces to the evaluation of various correlation functions 
%in the bosonic IKKT model, which can be done by 
%
%We perform explicit calculation
%up to the next-leading order, which reveals 
%
%Our results indeed reveal substantial effects of
We calculate
the internal energy, 
the Polyakov line,  and the extent of the
eigenvalue distribution
explicitly for 
%each case of 
$D=4,6,10$.
%Results at various $N$ show a clear scaling behavior,
%which allows us to extract the large-$N$ limit reliably.
%
%for $N=4,8,16,32$, and make large-$N$ extrapolations.
%and $D=4,6,10$.
%
%Comparison with the corresponding 
%results for purely bosonic theories
%indeed reveals substantial effects of fermions.
%,which are consistent with the behaviors suggested 
%by the gauge-gravity correspondence.
%
Our results nicely reproduce 
the asymptotic behaviors
of the recent Monte Carlo data obtained 
for both supersymmetric models and bosonic models
at finite temperature.
The different properties of
the two classes of models 
are clearly reflected
in the next-leading order terms.

The rest of this paper is organized as follows.
In section \ref{sec:model} we define the model
and the observables we study in this paper.
% and discuss their basic properties.
In sections \ref{section:DR} and \ref{section:nextb}
we present the calculations at the leading order
and at the next-leading order, respectively.
%In section \ref{section:Figure} we show explicit results
%for various quantities.
Section \ref{sec:Summary} is devoted to a summary and
discussions.
In Appendix \ref{sec:EO} we derive a formula,
which is used to calculate the internal energy.
In Appendix \ref{sec:Repre} we present the
form of Green's functions used to evaluate
them efficiently in actual Monte Carlo simulation.

%In practice, we can increase the statistics
%by exploitting the SO($D$) symmetry of the DR model
%as detailed in Appendix \ref{sec:Repre}. 
%
%
%In section \ref{sec:EO} we derive a formula
%for the internal energy, which is used for its
%evaluation at high temperature.

%In section \ref{sec:boundary} we study...

%==================================================================
\section{The models}
\label{sec:model}
%==================================================================
The models we study in this paper are
defined by the action
\beqa
\label{action}
S = \frac{1}{g^2} \intdt {\rm tr}
 \left\{
 \oot (D_t X_i)^2 
+\oot \psi_\alpha D_t \psi_\alpha
-\oof [X_i,X_j] ^2
- \oot \ps_\alpha (\gm_i)_{\alpha\beta} [X_i , \ps_\beta ]
\right\} \ ,
\eeqa  
where
$D_t \equiv \partial_t-i[A(t),\,\,\cdot\,\,]$ 
represents the covariant derivative.
The bosonic matrices
$A(t)$, $X_i(t)$ $(i=1,2,\cdots,d)$ 
and the fermionic matrices
$\ps_\al(t)$ $(\al=1,2,\cdots,p)$ are 
$N \times N$ Hermitian matrices,
where $p=4,8,16$ for $d=3,5,9$, respectively. 
The models can be obtained formally
by dimensionally reducing
${\cal N}=1$ super Yang-Mills theory
in $D=d+1$ dimensions to one dimension,
and they can be viewed as a 1d
gauge theory, where
$A(t)$, $X_i(t)$ and $\ps_\al(t)$
are the gauge field, adjoint scalars and 
spinors, respectively. 
The $p \times p$ symmetric matrices 
$\gm_i$ obey the Euclidean Clifford algebra
$\{ \gm_i,\gm_j \}=2\delta_{ij}$.
We impose 
periodic and anti-periodic boundary conditions
on the bosonic and fermionic matrices,
respectively.
The extent $\beta$ in the Euclidean time direction
$t$ then corresponds to the inverse
temperature $\beta=T^{-1}$.

The action is invariant under the shifts
\beq
A(t)  \mapsto  A(t) + \alpha(t) {\bf 1} \ , 
\quad
X_i(t)  \mapsto  X_i (t) + x_i {\bf 1} \ ,
\eeq
where $\alpha(t)$ is an arbitrary periodic function
and $x_i$ is an arbitrary constant.
In order to remove the corresponding decoupled modes,
we impose the conditions
\beqa 
\label{czm}
\tr  A(t) =0 \ , \quad
\intdt \tr  X_i (t) = 0 \quad (i=1,2,\cdots,d) \ .
\eeqa

The 't Hooft large-$N$ limit
corresponds to sending $N$ to $\infty$ with 
the 't Hooft coupling constant
$\lambda\equiv g^2 N$ fixed.
% , and the planar limit corresponds to 
%% Since $\lambda$ 
%% has a dimensionality of $(\rm{energy})^3$, 
%% the effective coupling constant $g^2_{\rm eff}$ is given by 
%% \beq
%% \label{g_eff}
%% g_{\rm eff}^2=\frac{\lambda}{T^3} \ .
%% \eeq
%% At high temperature,
%% non-zero frequency modes are weakly coupled,
%% and they can be integrated out perturbatively.
Since the coupling constant $g$ can be 
absorbed by rescaling the matrices and $t$
appropriately, we can set $\lambda$ to unity
without loss of generality.
%% \footnote{We are going to present 
%% explicit results for 
%% $8 \le N \le  32$, and the convention $\lambda\equiv 1$
%% is useful for extracting the planar limit.
%% However, one can extract results for 
%% other $N \rightarrow \infty$ limits if one wishes,
%% by rescaling our results appropriately.}
This implies that we replace the prefactor $\frac{1}{g^2}$
in the action (\ref{action}) by $N$ in what follows.

We define
%we calculate in the high temperature expansion 
the extent of the eigenvalue distribution 
and the Polyakov line
as
\beqa
%-------------------------------
\label{def R2}
R^2 &\equiv&
\frac{1}{N\beta} \intdt \tr \Bigl( X_i(t) \Bigr)^2 \ , 
\\
%-------------------------------
\label{def P}
P &\equiv& 
\frac{1}{N}\tr 
\mathcal{P} \exp
\left(
i\intdt A(t)
\right)
\ , 
\eeqa
where the symbol ``$\mathcal{P} \exp$'' 
represents the path-ordered exponential.

As a fundamental quantity in thermodynamics,
the free energy 
${\cal F} = - \frac{1}{\beta} \ln Z(\beta)$
is defined in terms of the partition function
given in the present model as
\beq
Z(\beta) = \int 
[{\cal D} A]_\beta
[{\cal D} X]_\beta  [{\cal D} \psi]_\beta 
 \, \ee^{- S(\beta)} \ ,
\label{def-part-fn}
\eeq
where the suffix of the measure $[ \ \cdot  \ ]_\beta $
represents the period of the field
to be path-integrated.
However, the evaluation of the partition function 
$Z(\beta)$ is not straightforward in 
Monte Carlo simulation, which we use 
for the integration over the zero modes.
%instead of some expectation values.
We therefore study the internal energy defined by
\beq
E \equiv \frac{d}{d \beta} (\beta {\cal F}) 
  = - \frac{d}{d \beta} \log Z(\beta) \ ,
\label{defE}
\eeq
which has equivalent information as 
the free energy,
given the boundary condition ${\cal F} = E$ at $T=0$.
Note also that the internal energy at $T=0$
provides the ground state energy of the quantum
mechanical system, which should vanish unless 
the supersymmetry is spontaneously broken.
In Appendix \ref{sec:EO}
we show that the internal
energy $E$
% (See eq.\ (\ref{defE}) for the definition)
can be expressed as
\beq
%-------------------------------
\label{def E}
\frac{E}{N^2} 
% - \frac{d}{d \beta} \log Z(\beta) 
= \langle {\cal E}_{\rm b}  \rangle + 
\langle {\cal E}_{\rm f} \rangle
%{\cal E}_{\rm b} +{\cal E}_{\rm f}
%
%\frac{3}{4}\, 
%\left(
%{\cal C}_1 +{\cal C}_2
%\right) 
\ ,
\eeq
where the operators 
${\cal E}_{\rm b}$ and ${\cal E}_{\rm f}$
are defined by
\beqa
%----------------------------------------------
\label{def-Eb}
{\cal E}_{\rm b} & \equiv &
%{\cal C}_1 
- \frac{3}{4}\, 
%\left\langle
\frac{1}{N\be}
\int_0^\beta  \!\!dt  \, \tr
\Bigl( [X_i,X_j]^2 \Bigr) 
%\right \rangle 
\ ,  \\
% \quad
%----------------------------------------------
{\cal E}_{\rm f} &\equiv& -
%{\cal C}_2 
\frac{3}{4}\, 
%\left\langle
\frac{1}{N\be} 
\int_0^\beta  \!\!dt  \, \tr
\Bigl( \ps_{\alpha} (\gm_i)_{\alpha\beta} [X_i,\ps_{\beta}] \Bigr)
%\right \rangle 
\  .
%----------------------------------------------
\label{def-Ef}
\eeqa
The symbol $\langle \ \cdot \ \rangle$
represents the expectation value 
with respect to the model (\ref{action}).

%% In this section we derive a formula
%% relating the internal energy of the present model
%% to some expectation value,
%% which can be calculated directly.
%% % Monte Carlo simulation.
%% %which is related to the first derivative of 
%% %the free energy with respect to the temperature.

%==================================================================
%\section{High temperature expansion}
%\label{sec:HTE}
%==================================================================

Let us take the static gauge $\partial_t A(t)=0$.
Correspondingly we add the ghost term
\beqa
\label{gfba} 
S_{\rm gh} 
= N \int_0^\beta \!\!dt \, \tr
\Bigl(
\partial_t
%\frac{d}{dt}
\overline{c}(t) D_t c(t)
\Bigr)
\eeqa 
to the action, where
$c$, $\bc$ are $N \times N$ matrices representing
the ghosts.
We make a Fourier expansion of the fields as
\beqa
&~& X_i(t) = \sum_n X_n^i \exp( i n  \omega t ) \ ,
\quad
\psi_\alpha (t)=\sum_r \psi_r^\alpha \exp(i r \omega t) \ , \\
&~& 
c(t)  = \sum_{n \neq 0} c_n  \exp(  i n \omega t ) \ , 
\quad
\bc(t)  = \sum_{n \neq 0} \bc _n  \exp( -  i n \omega t ) \ , 
\eeqa
where    
$\omega=\frac{2\pi}{\beta}$ represents the unit
of Matsubara frequencies, and the indices $n$ and $r$
% for the fermions
take integers and half-integers, respectively,
due to the imposed boundary conditions.
In terms of the Fourier modes, 
the gauge-fixed action is written as
\beqa
\label{base action}
\tilde{S} &=& S_{\rm 0}+S_{\rm kin}+S_{\rm int}
\ ,\\
%-----------------------------------------------------
\label{bIKKT action}
S_{\rm 0}
&\equiv& - 
N\be \ {\rm tr} 
\bigg\{ \frac{1}{2}\B([A, X_0^i]\B)^2
+ \frac{1}{4}\B([X_0^i,X_0^j]\B)^2 
\bigg\} \ ,
\\
%-----------------------------------------------------
\label{bkin action}
S_{\rm kin}
&\equiv&
N \be \ {\rm tr} 
\bigg\{ \oot \sum_{n \not=0} (n\omega)^2
X^i_{-n} X^i_{n}
+ \sum_{n \ne 0} (n\omega)^2 \bc_n c_n
+ \frac{1}{2} \sum_{r} i  r \omega \ps_{-r} \ps_{r} 
\bigg\} \ , \\
%=================================================
\label{bint action}
S_{\rm int}
&\equiv& -
N \be \ {\rm tr} 
\bigg\{
% *-*-*-*-*- cubic term -*-*-*-*-*
\sum_{n \ne 0}
n \omega
X^i_{-n} [A,X^i_{n}]
% --------------------------------
+ 
\sum_{n \not= 0}
n \omega
\bc_n [A,c_{n}]
%-----------------fermion----------
+ \frac{i}{2}
\sum_r \ps_{-r} [A,\ps_{r}]
%---------------------------
\\
% --------------------------------
% *-*-*-*-* quartic term *-*-*-*-*  
&&
+  \frac{1}{2}
\sum_{r,s} \ps_r \gm_i [X^i_{-r-s}, \ps_s]
+\oot\sum_{n \not= 0}
[A,X^i_{-n}][A,X^i_{n}]
% --------------------------------
+\oof
\sum_{npq}
%\!
{}' \,
[X^i_{-n-p-q},X^j_n][X^i_p,X^j_q]
% --------------------------------
\bigg\} \ , \nonumber
%---------
%\label{bint action}
%S_{\rm int}
%&\equiv& - \sum_{i=1}^6
%\mathcal{V}_i  \ .
%
\eeqa
where the symbol $\sum{}'$ implies 
that the $m\!=\! n\!=\! p\!=\! 0$ term is excluded.
%summing over indices with the constraint
%$m \!+\! n \!+\! p \!+\! q \!=\! 0$ excluding
%the case $m\!=\! n\!=\! p\!=\! q\!=\! 0$.

%we have omitted some terms from the quartic
%interactions, which will not be necessary for the next-leading
%calculation.

%===================================================================
\section{Leading order calculation}
\label{section:DR}
%===================================================================

In this section we consider 
the high temperature limit
\cite{Bal-Sathiapalan,Aharony:2004ig,Kawahara:2007nw},
which corresponds to the leading order calculation
at high $T$.
%the leading order calculation
%, which is discussed in refs.\ 
%contribution,
%which corresponds to taking the high temperature limit.
{}From (\ref{base action}),
one can easily see that all the non-zero modes decouple,
and one is left with the zero modes governed by the action 
(\ref{bIKKT action}).
By rescaling the zero modes as
\beqa
\label{gamma-alpha}
\tA_i \equiv T^{-1/4}X^i_0 \quad (i=1,2,\cdots, d) \ , 
\quad \tA_D \equiv T^{-1/4}A \ ,
% \quad (D \equiv d+1) \ ,
\eeqa
where we recall that $D \equiv d+1$,
the zero-mode action can be brought into the form
\beq
\label{DRmodel}
%S_{\rm DR} =
S_{0} =
\frac{1}{4} N \tr  (
\tilde{F}_{\mu\nu}
%[\tA_\mu,\tA_\nu]^2 
)^2 \ ,
\quad
\tilde{F}_{\mu\nu} = - i \,  [\tA_\mu,\tA_\nu] \ .
%-\frac{1}{4} N \tr \Big ([\tA_\mu,\tA_\nu]^2 \Big) \ .
\eeq
Here and henceforth, the Greek indices $\mu$, $\nu$ 
are assumed to run over $1,2, \cdots , D$.
The dimensionally reduced (DR) model (\ref{DRmodel})
is nothing but 
the bosonic part of the IKKT type matrix 
model \cite{Krauth:1998yu,HNT,Nishimura:2002va}.
%\footnote{The $d$ in the IKKT matrix model is $9$.} 
%``bosonic IKKT model'' below.
The leading behavior of the observables
at high temperature can be obtained as 
\beqa
%-----------------------------------------------------
\label{R2DR}
\left\langle R^2 \right\rangle &\simeq& 
T^{1/2}
%\sqT \ 
\left\langle \frac{1}{N} \,
\tr (\tA_i)^2 \right\rangle_{\rm DR} 
=
\chi_1 \, T^{1/2}
%\sqT
\ , \\
%-----------------------------------------------------
\label{PDR}
\left\langle P \right\rangle &\simeq&  
1 - \frac{1}{2} \, T^{-3/2}
\left\langle 
\frac{1}{N} \, \tr 
( \tA_D )^2
\right\rangle_{\rm DR} 
= 1-\frac{1}{2d} \, \chi_1 \,  T^{-3/2}
\ , \\
%-----------------------------------------------------
\label{F2DR}
\frac{1}{N^2} E
&\simeq& 
\frac{3}{4} \,  T \left\langle \frac{1}{N} \, 
\tr ( \tilde{F}_{ij}
%[\tA_i, \tA_j] 
)^2 \right\rangle_{\rm DR} 
=
\frac{3}{4} \, \chi_2 \,T \ , 
%-----------------------------------------------------
\eeqa
where $\langle \ \cdot \ \rangle_{\rm DR}$ 
represents the expectation value 
with respect to the DR model, 
and the coefficients $\chi_1$ and $\chi_2$ are given as
\beqa
%-----------------------------------------------------
\chi_1 &\equiv&
\left\langle \frac{1}{N} \ \tr (\tA_i)^2 \right\rangle_{{\rm DR}}
= \frac{d}{D}
\left\langle \frac{1}{N} \ \tr (\tA_\mu)^2 \right\rangle_{{\rm DR}}
%
%2.162(5)  
\label{DR1} \ , \\
%-----------------------------------------------------
\chi_2 &\equiv&
\left\langle \frac{1}{N} \
\tr (
% - [\tA_i, \tA_j]^2 
\tilde{F}_{ij}
)^2 \right\rangle_{{\rm DR}}
=  \frac{{}_d  {\rm C}_2}{{}_D {\rm C}_2}
\left\langle \frac{1}{N} \
\tr ( 
%- [\tA_\mu, \tA_\nu]^2 
\tilde{F}_{\mu\nu}
)^2 \right\rangle_{{\rm DR}}
=
(d-1) \left( 1-\frac{1}{N^2} \right) \ .
\label{DR2}
%----------------------------------------------------- 
\eeqa
We note that the expectation values
appearing here are standard quantities
calculated in the DR model (\ref{DRmodel})
by various methods \cite{HNT,Nishimura:2002va}.
In particular, the quantity in eq.\ (\ref{DR2}) can be
obtained exactly by simply rescaling the dynamical variables
or by writing down the Schwinger-Dyson equation \cite{HNT}.
%In order to calculate $\chi_1$ as well as other quantities
%appearing at the next-leading order in the next section,
%we perform Monte Carlo simulations.
%
%% The quantity in eq.\ (\ref{DR2}) can be
%% obtained exactly by simply rescaling the dynamical variables
%% or by writing down the Schwinger-Dyson equation \cite{HNT}.
%% In order to calculate $\chi_1$ as well as other quantities
%% appearing at the next-leading order in the next section,
%% we perform Monte Carlo simulations.
%% The results are summarized in table \ref{Tab_chi5678}. 
%% We have checked that the expectation value
%% appearing eq.\ (\ref{DR1}) agrees
%% with the previous results \cite{HNT}.
 
%==================================================================
\section{Next-leading order calculation}
\label{section:nextb}
%==================================================================

As is clear from the previous section, 
the leading order results are insensitive to the existence
of fermions.
In order to see their effects, we need to
proceed to the next-leading order calculation,
which involves the integration over the non-zero modes.

For that purpose, let us rescale the non-zero modes as 
\beq
\tX^i_n = \beta^{-1/2} \, X^i_n \ ,
\quad 
\tpsi_r = 
%(\beta \omega)^{1/2} \, 
\psi_r  \ , 
\quad
\tc_n = \beta^{-1/2}  \, \tc_n \ ,
\quad
\tbc_n = \beta^{-1/2} \, \tbc_n  \ , 
\eeq
where $n\neq 0$, 
so that the kinetic terms take the canonical form
\beq
\label{bkin action2}
S_{\rm kin}
\equiv
N \ {\rm tr} 
\bigg\{ \oot \sum_{n \not=0} (2 \pi n)^2
\, \tX^i_{-n} \tX^i_{n}
+ \sum_{n \ne 0} (2\pi n)^2 \, \tbc_n \tc_n
+ \frac{1}{2} \sum_{r} 2 \pi i  r \, \tpsi_{-r} \tpsi_{r} 
\bigg\} \ . \\
\eeq
Then the propagators are given by
\beqa
%-----------------------------------------------
\label{pro X}
\B\langle\!\!\B\langle
(\tX^i_m)_{ab}(\tX^j_n)_{cd}
\B\rangle\!\!\B\rangle
&=&
\frac{1}{(2 \pi n)^2 N } \
\dl_{ij} \dl_{m,-n} \dl_{ad} \dl_{bc} \ ,
\\
%-----------------------------------------------
\label{pro psi}
\B\langle\!\!\B\langle
(\tpsi_{\al r})_{ab}(\tpsi_{\be s})_{cd}
\B\rangle\!\!\B\rangle
&=&
\frac{1}{ 2 \pi i  r N} \
\dl_{\al \be} \dl_{r,-s} \dl_{ad} \dl_{bc} \ , \\
%--------------------
\label{pro c}
\B\langle\!\!\B\langle
(\tc_m)_{ab}(\tbc_n)_{cd}
\B\rangle\!\!\B\rangle
&=&
\frac{1}{ (2 \pi n) ^2 N} \
\dl_{mn} \dl_{ad} \dl_{bc} \ ,
%-----------------------------------------------
\eeqa
where the symbol $\langle \! \langle \,\,\, 
\cdot \,\,\,\rangle \!\rangle$ represents
integrating {\em only the non-zero modes}
using the quadratic terms (\ref{bkin action2}).
%$S_{\rm kin}$.
The interaction terms are given by
$S_{\rm int} \equiv - \sum_{i=1}^6 \mathcal{V}_i$, where
\beqa
\label{V}
%----------------------------------
\mathcal{V}_1 &\equiv&
\sqrt{\epsilon} N 
\sum_{n \ne 0}
n \,  \tr
\B(\tX^i_{-n}[\tA_D, \tX^i_{n}]\B)
 \ , \quad
%----------------------------------
\mathcal{V}_2 \equiv
\sqrt{\epsilon} N 
\sum_{n \not= 0}
n \, \tr
\B(\tbc_{n}[\tA_D, \tc_{n}]\B)
 \ ,\nonumber\\
%----------------------------------
\mathcal{V}_3 
&\equiv&
\frac{i}{2} \sqrt{\epsilon} N \sum_{r} \tr
\B(\tpsi_{-r} [\tA_D ,\tpsi_{r}]\B)
\ , \quad
%----------------------------------
\mathcal{V}_4 
\equiv 
\frac{1}{2} \sqrt{\epsilon} N \sum_{r} \tr
\B(\tpsi_{-r} \gm_i [ \tA_i,\tpsi_{r}]\B)
\ , \nonumber \\
%-----------
\mathcal{V}_5 &\equiv&
\frac{1}{2} \, \epsilon  N \sum_{n \not=0} \tr
\B([\tA_D, \tX^i_{-n}][\tA_D, \tX^i_{n}]\B)
\ ,
\nonumber\\
%----------------------------------
\mathcal{V}_6 
&\equiv&
\frac{1}{2} \, \epsilon  N   \sum_{n \not= 0} 
\tr \B( 
[\tA_i,\tX^j_{-n}][\tA_i,\tX^j_{n}] 
+ [\tA_i, \tX^j_{-n}][\tX^i_{n},\tA_j ]
%+ [\tA_i,\tA_j ][\tX^i_{-n},\tX^j_{n} ]
 \B) \ .
%\nonumber\\
%\oof \epsilon \sum{}' \,
%[X^i_m,X^j_n][X^i_p,X^j_q] \ ,
%\nonumber  \\
% --------------------------------
%% \mathcal{V}_4 
%% &\equiv&
%% \frac{1}{2} N\be \sum_{n \not= 0} 
%% \tr \B( 
%% [X^i_0,X^j_n][X^i_0,X^j_{-n}] 
%% +
%% [X^i_0,X^j_n][X^j_{-n},X^i_0] \B) \ ,
%% \nonumber\\
%----------------------------------
\eeqa
We have introduced the expansion parameter\footnote{If we
left the 't Hooft coupling constant
$\lambda$
%$ = g^2 N$ 
arbitrary, we would find that
the expansion parameter is given by $\epsilon  \sqrt{\lambda}
= \sqrt{\lambda/T^3}$. This is what one might have
deduced on dimensional grounds, since the 't Hooft coupling
constant has the dimension of $({\rm mass})^3$ 
in the present models.}
$\epsilon = \beta^{3/2}$,
%\beq
%\epsilon = \frac{\beta^{3/2}}{(2\pi)^2} \ ,
%\eeq
%and in eq.\ (\ref{V}) 
and omitted terms,
%higher terms in $\epsilon$,
which are irrelevant to the calculations
at the next-leading order.

First we calculate the extent of the eigenvalue distribution
(\ref{def R2}), which can be decomposed as
\beqa
\label{R2 mode}
R^2 
%= \frac{1}{N \be}
%\int dt \, \tr (X_i)^2 
= \ooN
 \tr (X_0^i)^2 + \ooN \sum_{n \not= 0}
 \tr (X_n^i X_{-n}^i)  \ .
\eeqa
Let us consider the first term.
The leading order contribution is given by
(\ref{R2DR}).
At the next-leading order, 
we use the vertices (\ref{V}) and 
integrate over the non-zero modes at one-loop
making use of the formulae
\beq
\sum_{n\neq 0} \frac{1}{(2\pi n)^2} = \frac{1}{12} \ ,
\quad
\sum_{r} \frac{1}{(2\pi r)^2} = \frac{1}{4} 
\eeq
to sum over the Matsubara frequencies in the loop.
This gives rise to the operators written in terms
of zero modes as
\beqa
%----------------------------------------------------
\mathcal{O}_1
&\equiv&
\frac{1}{2} 
\langle\!\langle \,
%\B\langle\!\!\B\langle 
(\mathcal{V}_1)^2 
\, \rangle\!\rangle 
%\B\rangle\!\!\B\rangle
= \frac{d}{6} N \be^{3/2} \tr (\tA_D)^2
%-\bigg( \ooN \tr A \bigg)^2
 \ ,
\quad
%----------------------------------------------------
\mathcal{O}_2
\equiv 
\frac{1}{2} 
\langle\!\langle \,
%\B\langle\!\!\B\langle
(\mathcal{V}_2 )^2
\,\rangle\!\rangle 
%\B\rangle\!\!\B\rangle
= - \frac{1}{12} N \be^{3/2} \tr (\tA_D)^2
%= -\frac{1}{12}(N \be)^2 
%\left(
%\cdot \ooN \tr A{}^2
%+\bigg( \ooN \tr A \bigg)^2
%\right)
\ ,\\
%----------------------------------------------------
%----------------------------------------------------
\mathcal{O}_3
&\equiv& 
\frac{1}{2}
\langle\!\langle \,
%\B\langle\!\!\B\langle 
(\mathcal{V}_3)^2
\, \rangle\!\rangle 
%\B\rangle\!\!\B\rangle 
= - \frac{p}{8} N \be^{3/2} \tr (\tA_D)^2
%= -\frac{p}{8} N \be
%\left(
%\cdot \frac{1}{N} \tr (A_0)^2
%+\bigg( \ooN \tr A \bigg)^2
%\right)
 \ ,
\quad
%----------------------------------------------------
\mathcal{O}_4
\equiv 
\frac{1}{2}
\langle\!\langle \,
%\B\langle\!\!\B\langle 
(\mathcal{V}_4)^2
\, \rangle\!\rangle 
%\B\rangle\!\!\B\rangle 
= \frac{p}{8} N \be^{3/2} \tr (\tA_i)^2
%= \frac{p}{8} N \be
%\cdot \frac{1}{N}  \tr (X_0^i)^2
%-\bigg( \ooN \tr X_i \bigg)^2  
\ ,\\
%----------------------------------------------------
\mathcal{O}_5
&\equiv& 
\langle\!\langle \,
\mathcal{V}_5 \,
\rangle\!\rangle
= - \frac{d}{12} N \be^{3/2} \tr (\tA_D)^2
%= -\frac{d}{12} (N \be)^2  
%\left(
%\cdot \ooN \tr A{}^2 
%+\bigg( \ooN \tr A \bigg)^2
%\right)
\ ,
\quad
%----------------------------------------------------
\label{K4}
\mathcal{O}_6
\equiv 
\langle\!\langle \,
\mathcal{V}_6  \, 
\rangle\!\rangle
= - \frac{d-1}{12} N \be^{3/2} \tr (\tA_i)^2
% = \frac{1}{12} (1-d)(N \be)^2  
%\left(
%\cdot \ooN \tr (X_0^i)^2
%-\bigg( \ooN \tr X_0^i \bigg)^2
%\right)
\ .
\eeqa
Summing up these operators, we obtain
\beq
\mathcal{O} = 
\sum_{j=1}^6  \mathcal{O}_j
= - \Big( \frac{d-1}{12} - \frac{p}{8}
\Big)  N \be^{3/2} \B\{  \tr (\tA_i)^2 
-  \tr (\tA_D)^2 \B\} \  .
\eeq
Using this operator, we can evaluate the first term 
in eq.\ (\ref{R2 mode}) as
\beqa
\label{1_1-loop_R2}
\ooN
\B\langle
\tr (X_0^i)^2 \B\rangle &\simeq&
% _{\rm NLO} =
\left\langle
\ooN \ \tr (X_0^i)^2 
\right\rangle_{\rm DR}
+ \left\langle
\ooN \ \tr (X_0^i)^2 \,\cdot\,
 \mathcal{O}
\right\rangle_{\rm DR,c} \\
&=& \chi_1 \, T^{1/2} 
- \Big( \frac{d-1}{12} - \frac{p}{8}
\Big) (\chi_{\rm 3}-\chi_{\rm 4}) \, T^{-1} 
+ {\rm O}(T^{-5/2})
\ ,
% + \frac{1}{4}\B( \frac{1-d}{3}+\frac{p}{2} \B)
%(\chi_{\rm 3}-\chi_{\rm 4}) \, T^{-1} \ ,
\eeqa
where the subscript ``c'' implies
that the connected part is taken,
and we define the coefficients $\chi_3$ and $\chi_4$ by
\beqa
%--------------------------------------------
\chi_3 \equiv  
\B\langle
\tr (\tA_i)^2 \cdot \tr (\tA_j)^2 
\B\rangle_{\rm DR,c}\ ,\quad
%--------------------------------------------
\chi_4 \equiv 
\B\langle
\tr (\tA_i)^2 \cdot \tr (\tA_D)^2 
\B\rangle_{\rm DR,c} \ .
%--------------------------------------------
\label{chi3-4}
\eeqa
The second term of eq.\ (\ref{R2 mode})
can be calculated at the next-leading order 
using the propagator (\ref{pro X}), and we get 
\beqa
\label{bs0}
\ooN \sum_{n \not=0} \B\langle \tr (X^i_n X^i_{-n}) \B\rangle
\simeq 
\ooN \sum_{n \not=0} \B\langle\!\!\B\langle \tr (X^i_n X^i_{-n}) 
\B\rangle\!\!\B\rangle
=
\frac{d}{12}\, T^{-1}+ {\rm O}(T^{-5/2}) \ .
\eeqa
Adding the two terms, we get the result at the next-leading
order as 
\beqa 
\label{HTE_bR2}
\big\langle
R^2 
\big\rangle
&= &     
\chi_1 T^{1/2}
+ \left\{ \frac{d}{12}
- \Big( \frac{d-1}{12} - \frac{p}{8}
\Big) (\chi_{\rm 3}-\chi_{\rm 4}) \right\}
T^{-1}
+ {\rm O}(T^{-5/2}) \ .%&=&
\eeqa

Let us calculate the internal energy
(\ref{defE}) using (\ref{def E}).
The operators ${\cal E}_{\rm b}$
and ${\cal E}_{\rm f}$ can be decomposed as
\beq
{\cal E}_{\rm b}
= \frac{3T}{4N}
% T \frac{1}{N}
\tr (\tilde{F}_{ij})^2
- \frac{3}{N^2\beta} {\cal V}_6 +  \cdots \ , 
\quad
{\cal E}_{\rm f}
= - \frac{3}{2 N^2\beta} {\cal V}_4  + \cdots \ ,
\eeq
where we have omitted terms irrelevant
at the next-leading order.
The expectation values can be calculated as
\beqa
%-------------------------------------------------
\label{HTE_E}
\langle {\cal E}_{\rm b} \rangle 
&\simeq&
\frac{3}{4} T
\left\{ \left\langle 
\frac{1}{N} \, \tr 
( \tilde{F}_{ij} )^2 \right\rangle_{\rm DR} 
+ \left\langle 
\frac{1}{N} \, \tr 
( \tilde{F}_{ij} )^2 \cdot {\cal O} \right\rangle_{\rm DR,c}  \right\} 
- \frac{3}{N^2 \beta} 
\Bigl\langle  
\langle\!\langle \,
{\cal V}_6  \,
\rangle\!\rangle 
\Bigr\rangle_{\rm DR}  \\
&=&
\frac{3}{4} \chi_2 \, T
+ \left\{ 
\frac{1}{4}(d-1) \chi_1
- \frac{3}{4}
\B( \frac{d-1}{12}-\frac{p}{8} \B)
(\chi_5 - \chi_6 )
 \right\} T^{-1/2}
 + {\rm O} \big(T^{-2}\big)  \ , \\
\langle {\cal E}_{\rm f} \rangle 
&\simeq&
- \frac{3}{2 N^2 \beta} 
\Bigl \langle  
\langle\!\langle \,
({\cal V}_4)^2  \,
\rangle\!\rangle 
\Bigr\rangle_{\rm DR}
= 
-\frac{3p}{8} \,\chi_1 \, T^{-1/2}
+ {\rm O} \big(T^{-2}\big) \ ,
\eeqa
where we define the coefficients
\beq
\chi_5 \equiv 
\B\langle 
\tr 
( \tilde{F}_{ij} )^2
%- [\tA_i, \tA_j]^2 
\cdot
\tr (\tA_k)^2 
\B\rangle_{\rm DR,c} \ , 
\quad
%---------------------------------------------
\chi_6 \equiv 
\B\langle 
\tr ( \tilde{F}_{ij} )^2
% - [\tA_i, \tA_j]^2  
\cdot 
\tr (\tA_D)^2 
\B\rangle_{\rm DR,c} \ .
\label{chi5-6}
\eeq
Adding these terms, we get the result for the internal
energy as
\beq
\frac{E}{N^2} 
= \frac{3}{4}\, \chi_2 \, T
%-\frac{1}{32}\B\{ 2(d-1)-3p \B\}
- \frac{3}{4}
\B( \frac{d-1}{12}-\frac{p}{8} \B)
(\chi_5 - \chi_6 - 4\chi_1)
\, T^{-1/2}
%\left\{ 
%(\chi_5-\chi_6)
%-4\B(2(d-1)-3p \B)\,\chi_1
%\right\}
+ {\rm O} \big(T^{-2}\big)
\ .
%\ .\nonumber\\
%-------------------------------------------------
\label{HTE_P}
\eeq

Similarly the Polyakov line can be calculated as
\beqa
\langle P \rangle &\simeq&
1 - \frac{1}{2} \, T^{-3/2}
\left\{ \left\langle 
\frac{1}{N} \, \tr 
( \tA_D )^2 \right\rangle_{\rm DR} 
+ \left\langle 
\frac{1}{N} \, \tr 
( \tA_D )^2 \cdot {\cal O} \right\rangle_{\rm DR,c}  \right\} 
\nonumber \\
&~& + \frac{1}{24} \, T^{-3}
\left\langle 
\frac{1}{N} \, \tr 
( \tA_D )^4
\right\rangle_{\rm DR}  
\nonumber \\
&=&
1-\frac{1}{2d}\,\chi_1 T^{-3/2}
+ \left\{\frac{1}{24} \chi_8
+ \frac{1}{2}
\B( \frac{d-1}{12}-\frac{p}{8} \B)(\chi_4-\chi_7)
\right\} T^{-3}
+ {\rm O}\big(T^{-9/2}\big)\ , 
\label{Pol_res}
%\nonumber
%-----------------------------------------------------
\eeqa
where we define the coefficients
$\chi_7$ and $\chi_8$ by
\beq
\chi_7 \equiv
\B\langle 
\tr (\tA_D)^2 
\cdot 
\tr (\tA_D)^2 
\B\rangle_{\rm DR,c} \ , 
\quad 
\chi_8 \equiv  
\B\langle 
\frac{1}{N}\,\tr (\tA_D)^4 
\B\rangle_{\rm DR} \ .
\label{chi7-8}
\eeq
%where $F_{ij} \equiv i\,[X_0^i,X_0^j]$.

Thus we have obtained various quantities
up to the next-leading order with
the coefficients $\chi_i$, which can be 
obtained by Monte Carlo evaluation of
the connected Green's functions
in the DR model (\ref{DRmodel}).
%using  simulation.
In practice, we rewrite the Green's functions
as described in Appendix \ref{sec:Repre}
in order to increase the statistics.
The values of $\chi_i$ 
obtained in this way\footnote{The heat-bath algorithm as
described in ref.\ \cite{HNT} has been used for
Monte Carlo simulation.
We have made 25M sweeps for $d=3,5$ 
and 8M sweeps for $d=9$ to obtain the data.
We have checked that the expectation value
appearing in the definition (\ref{DR1}) 
of $\chi_1$ agrees
with the previous results \cite{HNT}.
}
for $d=3,5,9$ and for various $N$
are summarized in table \ref{Tab_chi5678}.
In order to see the large-$N$ behavior \cite{HNT},
we plot the values of $\chi_i$
against $1/N^2$ for $12 \le N \le 32$ 
in fig.\ \ref{Fig_HTE data}.
We observe that 
the data points for $N=16,20,32$ lie on
a straight line.
This enables us to obtain
the large-$N$ extrapolated values 
shown in table \ref{Tab_chi5678}.
%table \ref{Tab_chi5678}.
%\section{Results}
%\label{section:Figure}   

Using the coefficients $\chi_i$ extrapolated to $N=\infty$,
we evaluate the expressions (\ref{HTE_bR2}),
(\ref{HTE_P}) and (\ref{Pol_res}).
%as a function of $T$.
The results for the bosonic case
can be readily obtained by setting $p=0$
in the same expressions.
In fig.\ \ref{Fig_HTE d3} we show 
various quantities as a function of $T$
for $d=3$ (left column) and $d=9$ (right column), 
respectively.
The curves represent the results of
the high temperature expansion\footnote{In refs.\
\cite{Kawahara:2007nw,Kawahara:2007fn,Anagnostopoulos:2007fw},
we have presented the results of the high temperature
expansion using the coefficients $\chi_i$ obtained at 
the same $N$ as those used for Monte Carlo simulations
at finite temperature.
The quality of the agreement with the Monte Carlo data
at high $T$ is almost the same as in the present plots.}.
The solid lines represent
the leading order results, which 
are the same for 
%do not distinguish
%(They are the same
%to 
%for 
the bosonic and supersymmetric cases.
The dashed lines and the dash-dotted lines
represent the next-leading order results
for the bosonic case
and the supersymmetric case, respectively.
For comparison, we also plot the recent Monte Carlo data
obtained at finite $T$
for the bosonic model with $d=3$ 
\cite{Kawahara:2007nw}
and $d=9$ \cite{Kawahara:2007fn},
and for the supersymmetric model with $d=9$
\cite{Anagnostopoulos:2007fw}.
In both bosonic and supersymmetric cases,
the high temperature expansion including the next-leading
order terms seems to be valid at $T \gtrsim 2$.

%In both figures the results
%for $N = 4$ and $N=16$ are shown in the left and 
%right columns, respectively.

%% \TABLE[pos]{\small%
%% \begin{tabular}{|c|c|c|c|c|c|c|c|c|}
%% \hline 
%% {$d$} & {$\chi_1$} & {$\chi_3$} & {$\chi_4$} & {$\chi_5$} & 
%% {$\chi_6$} & {$\chi_7$} & {$\chi_8$}  \\
%% \hline \hline 
%% {3} &  {1.6150(1)}   & {1.83(1)}  & {-0.6039(6)} 
%% & {2.676(9)} & {-0.940(2)}  & {1.016(5)}  & {0.6051(2)}  \\
%% \hline 
%% {5} &  {1.8382(8)} & {1.01(1)}  & {-0.229(3)}  
%% & {3.27(4)}  & {-0.751(9)} & {0.384(1)}  & {0.27868(5)}  \\
%% \hline 
%% {9} &  {2.2975(1)}  & {0.719(6)} & {-0.07(1)}  & {4.29(2)} 
%% &  {-0.59(2)} & {0.14(2)} & {0.13257(5)} \\
%% \hline
%% \end{tabular} 
%% \caption{
%% The results for $\chi_i \ (i=1,3,\cdots,8)$ in various $d$ 
%% obtained from extrapolating the results 
%% for various finite $N$ as shown in fig.\ref{Fig_HTE data}.
%% \label{Tab_chi5678}  
%% } 
%% }

%\pagebreak[4]
%\newpage

\FIGURE{
\epsfig{file=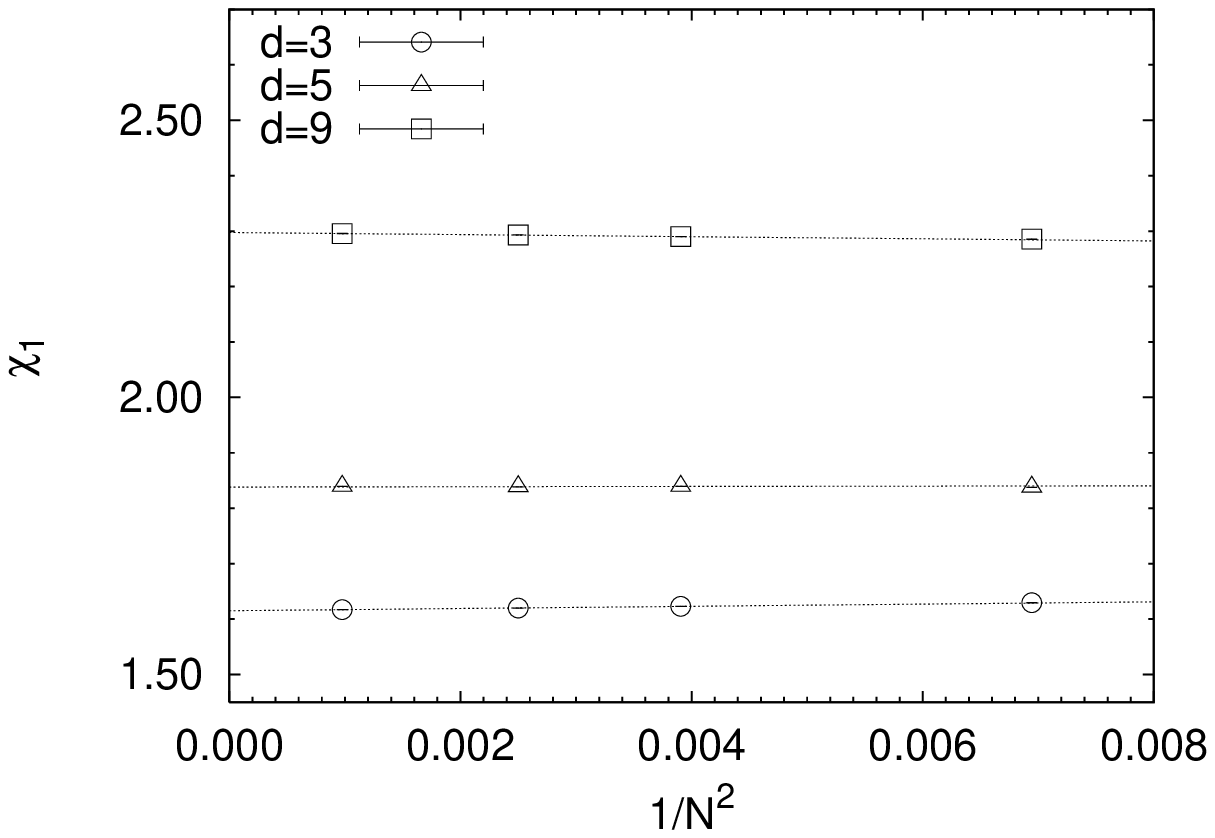,width=7.0cm} 
\epsfig{file=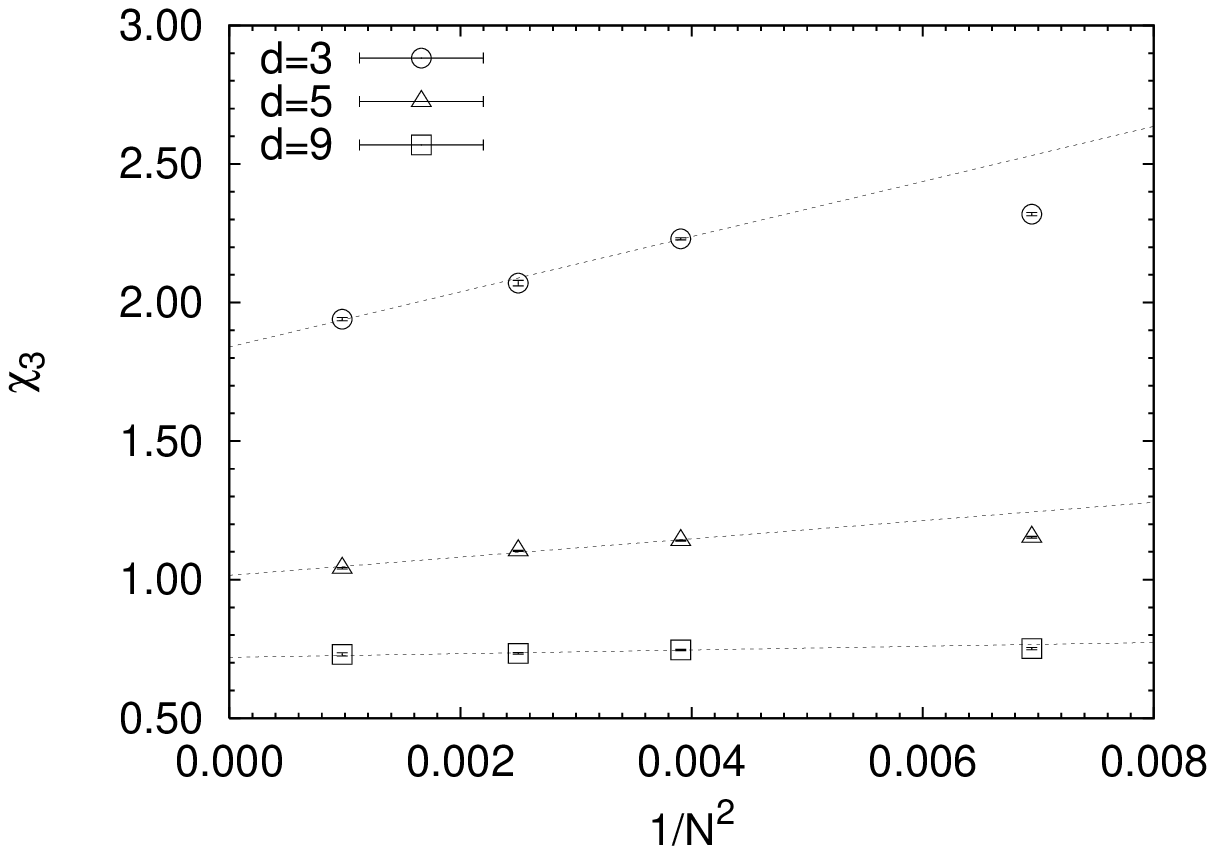,width=7.0cm}
\epsfig{file=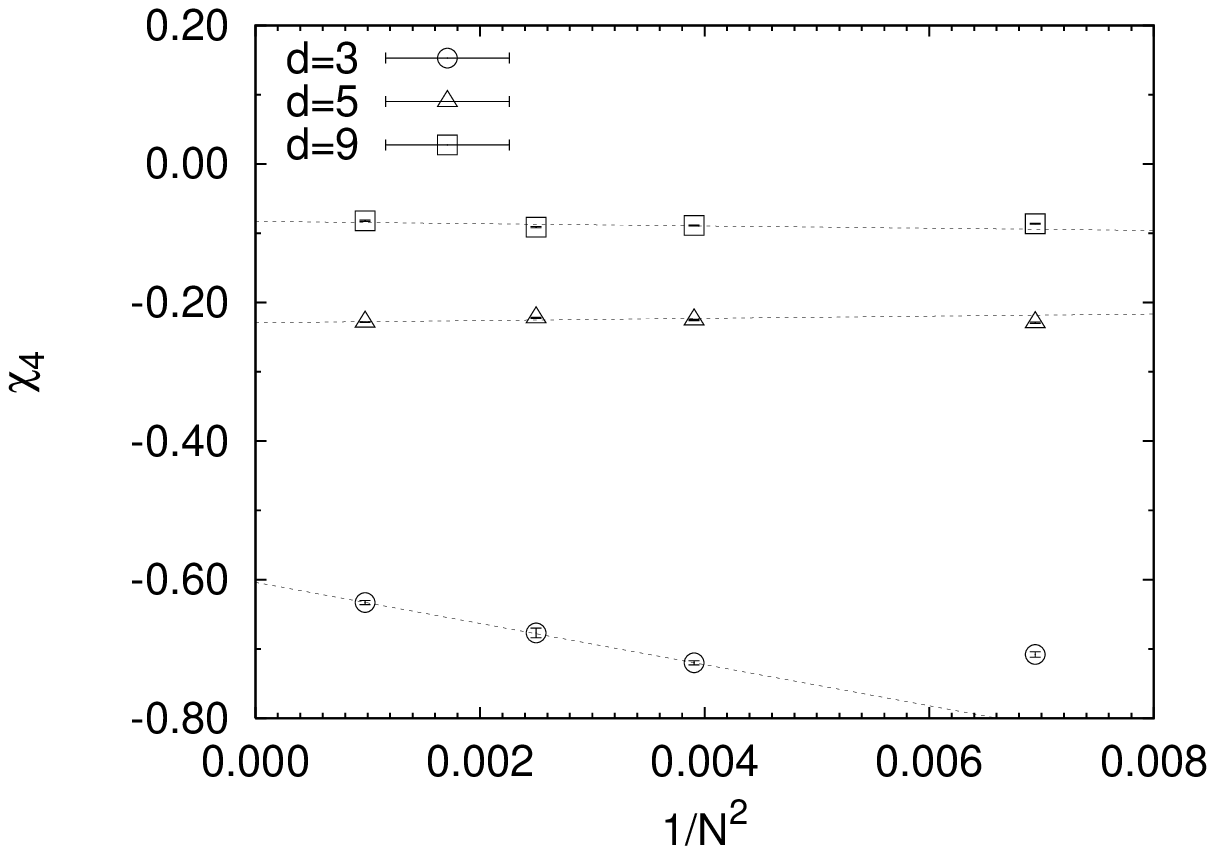,width=7.0cm}
\epsfig{file=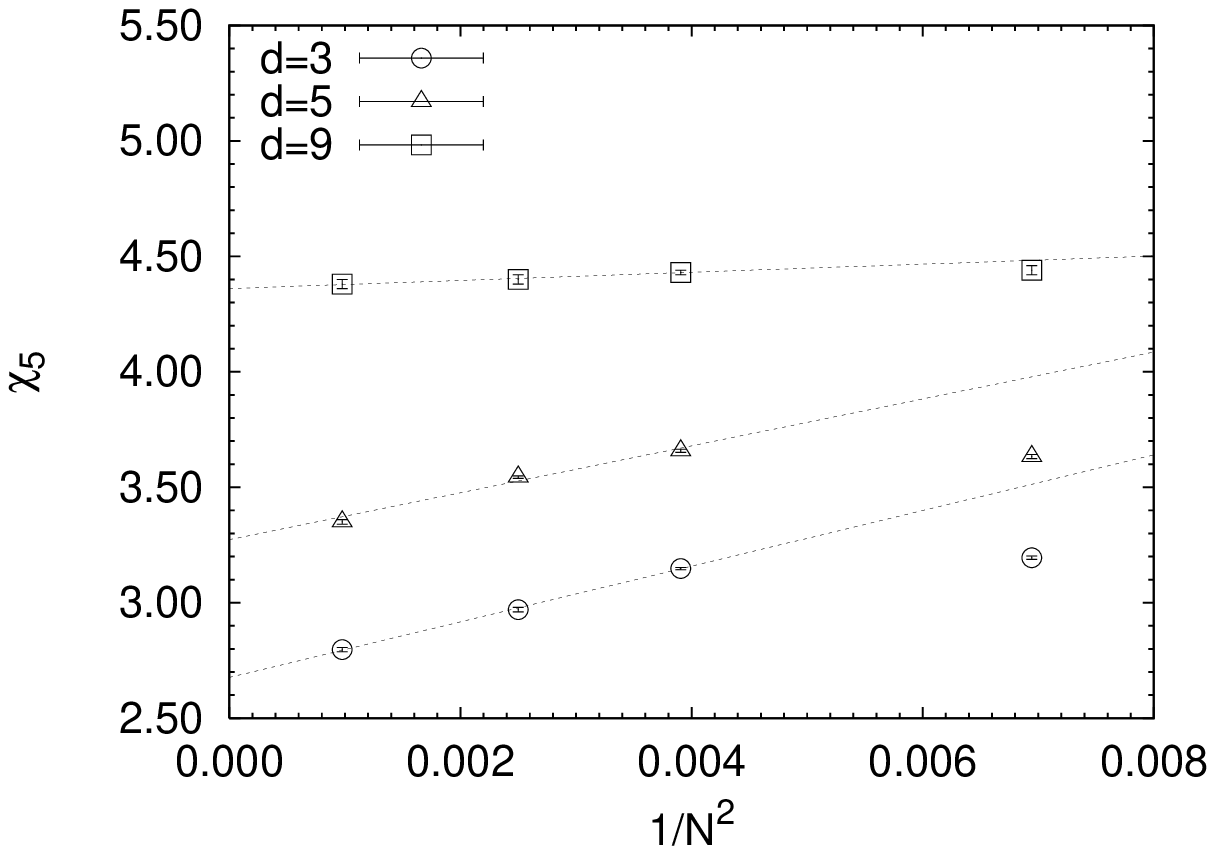,width=7.0cm} 
\epsfig{file=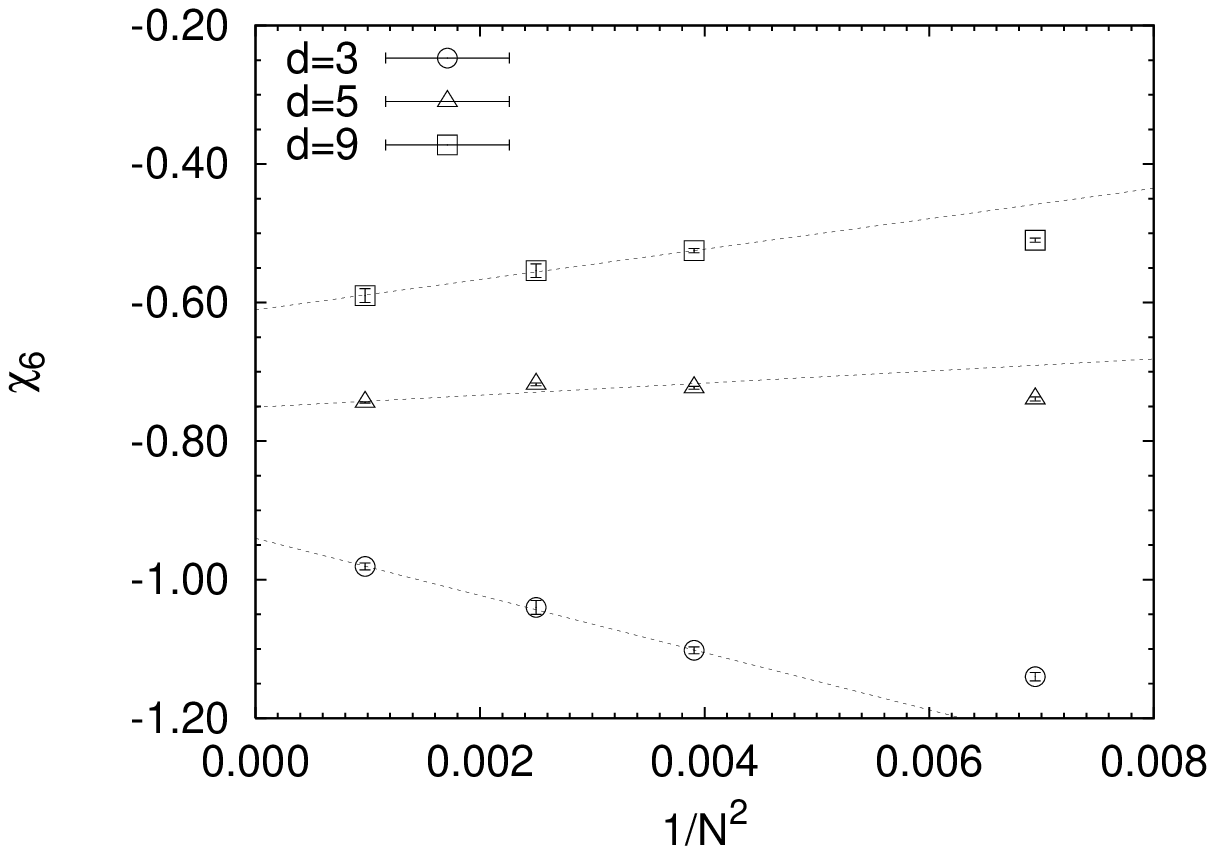,width=7.0cm}  
\epsfig{file=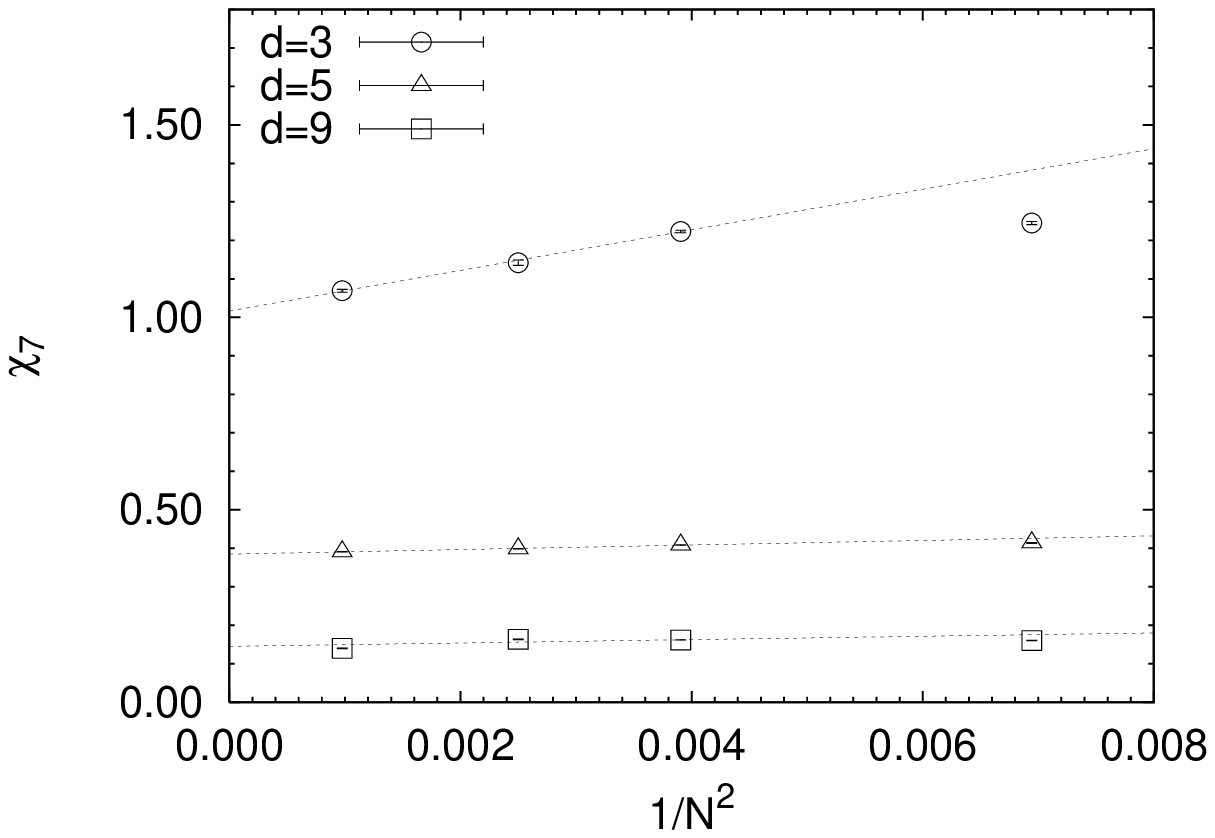,width=7.0cm} 
\epsfig{file=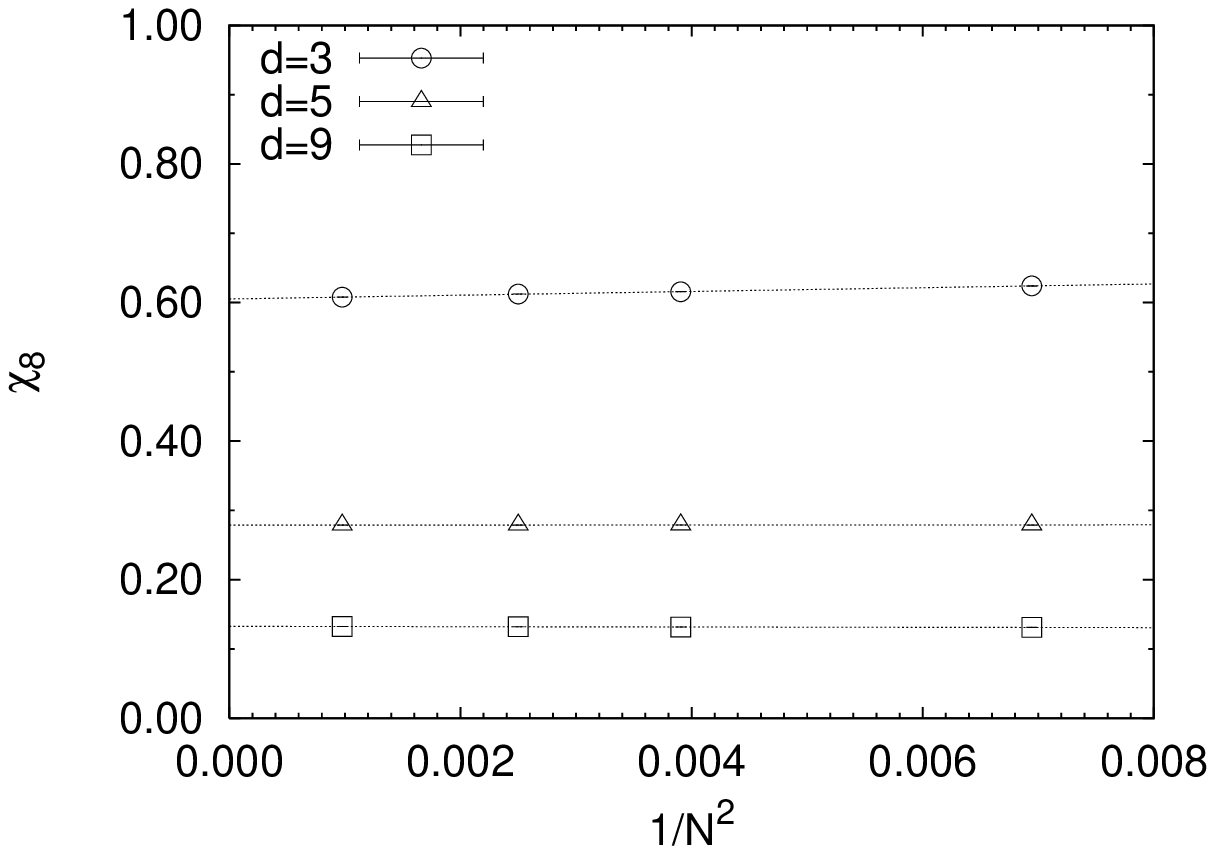,width=7.0cm}
\caption{The coefficients $\chi_i \ (i=1,3,\cdots,8)$
for $d=3,5,9$ and $12 \le N \le 32$
evaluated by Monte Carlo simulation of the 
corresponding DR model are plotted against $1/N^2$. 
The straight lines represent fits to
the expected large-$N$ behavior $a+b/N^2$
using the $N=16,20,32$ data points.
The extrapolated values 
are shown in table \ref{Tab_chi5678}
as results at $N=\infty$.  
}     
\label{Fig_HTE data}} 

%\newpage
%\pagebreak[4]

\FIGURE{
\epsfig{file=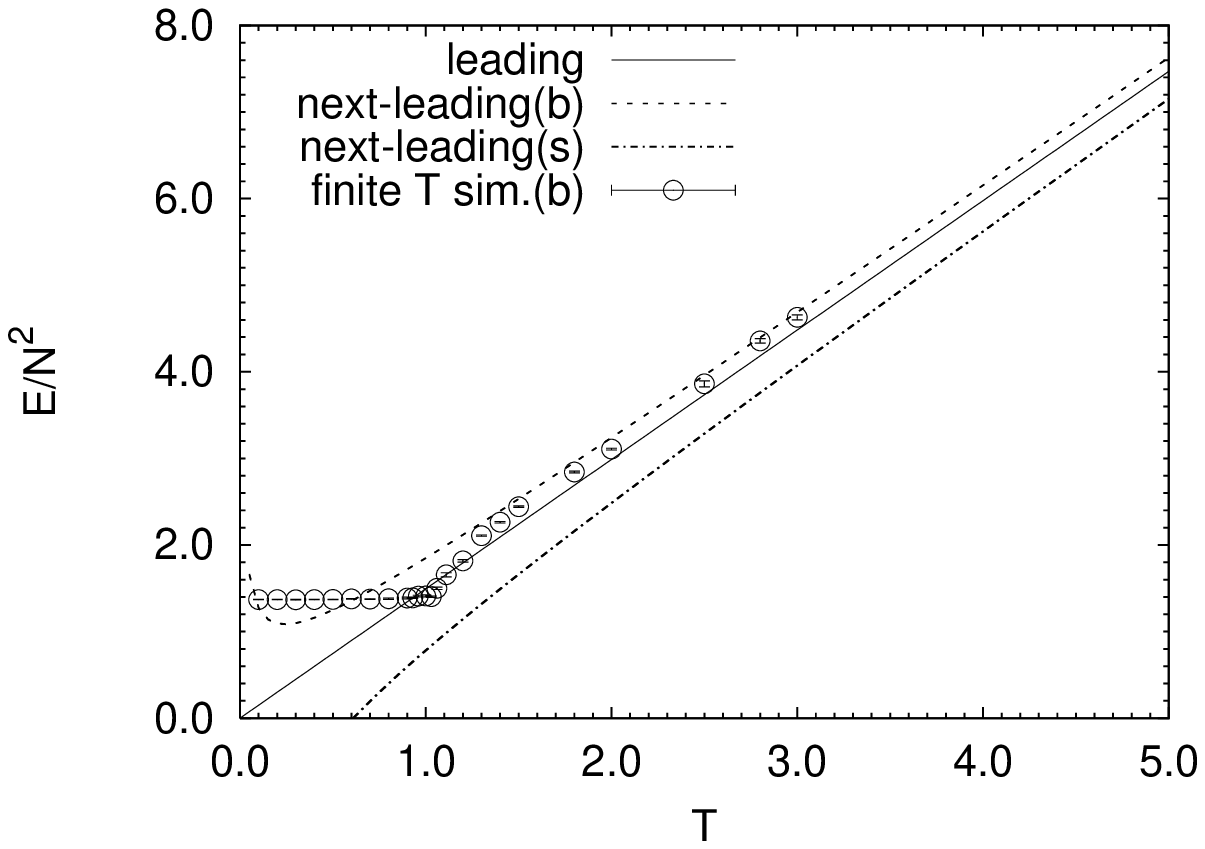,width=7.0cm} 
\epsfig{file=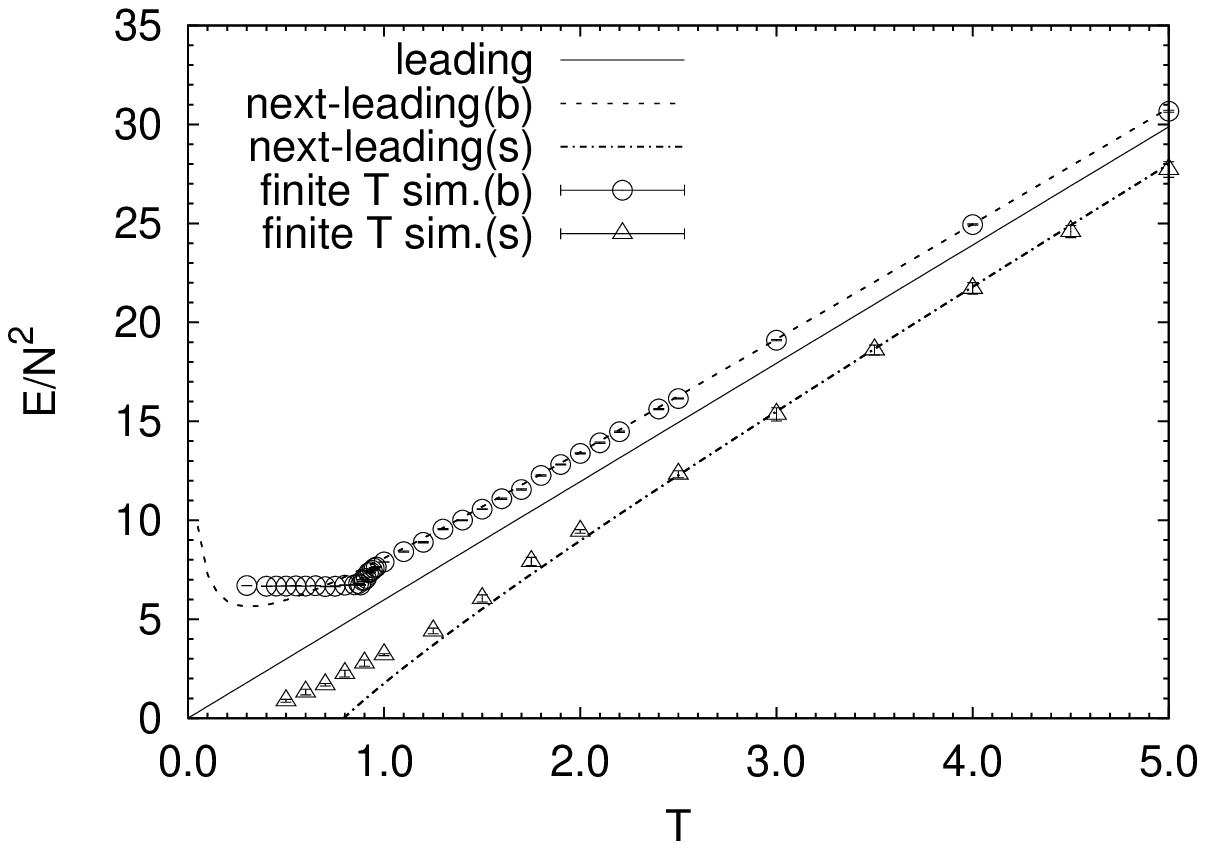,width=7.0cm}
\epsfig{file=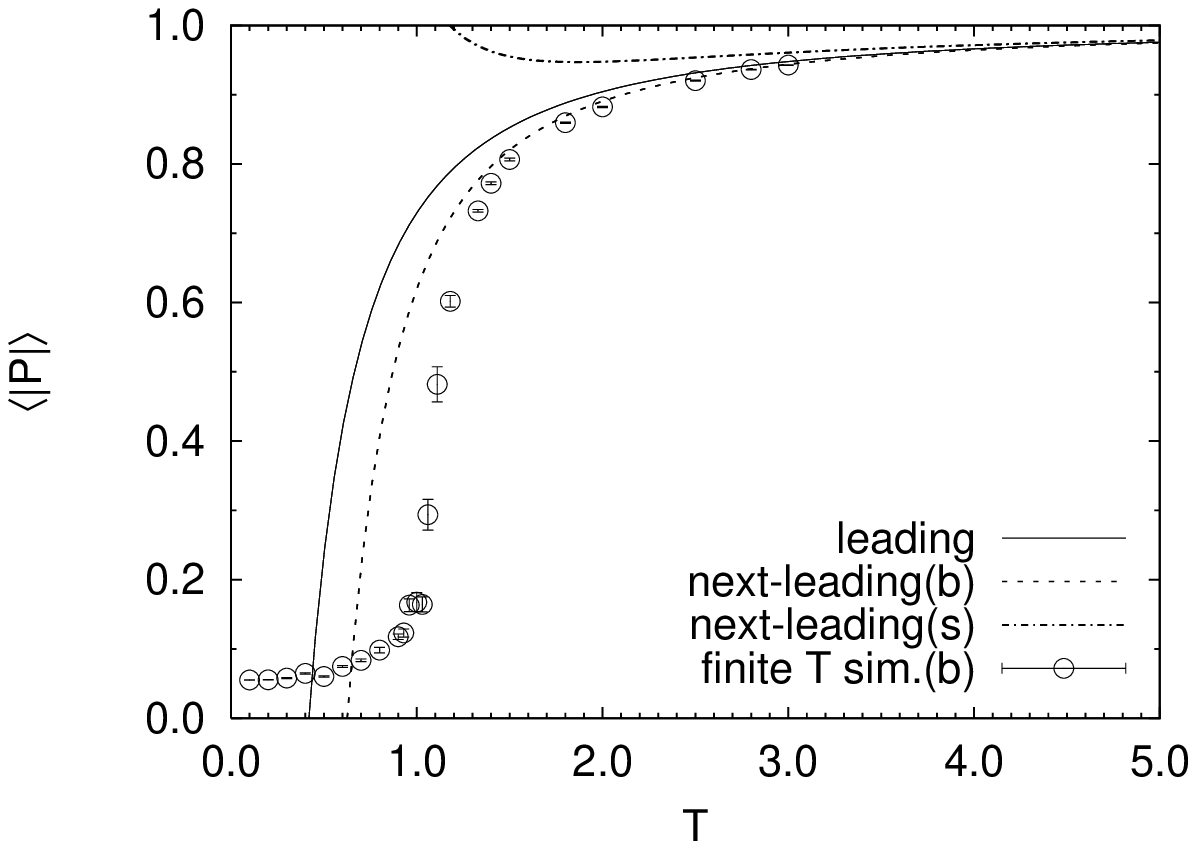,width=7.0cm}
\epsfig{file=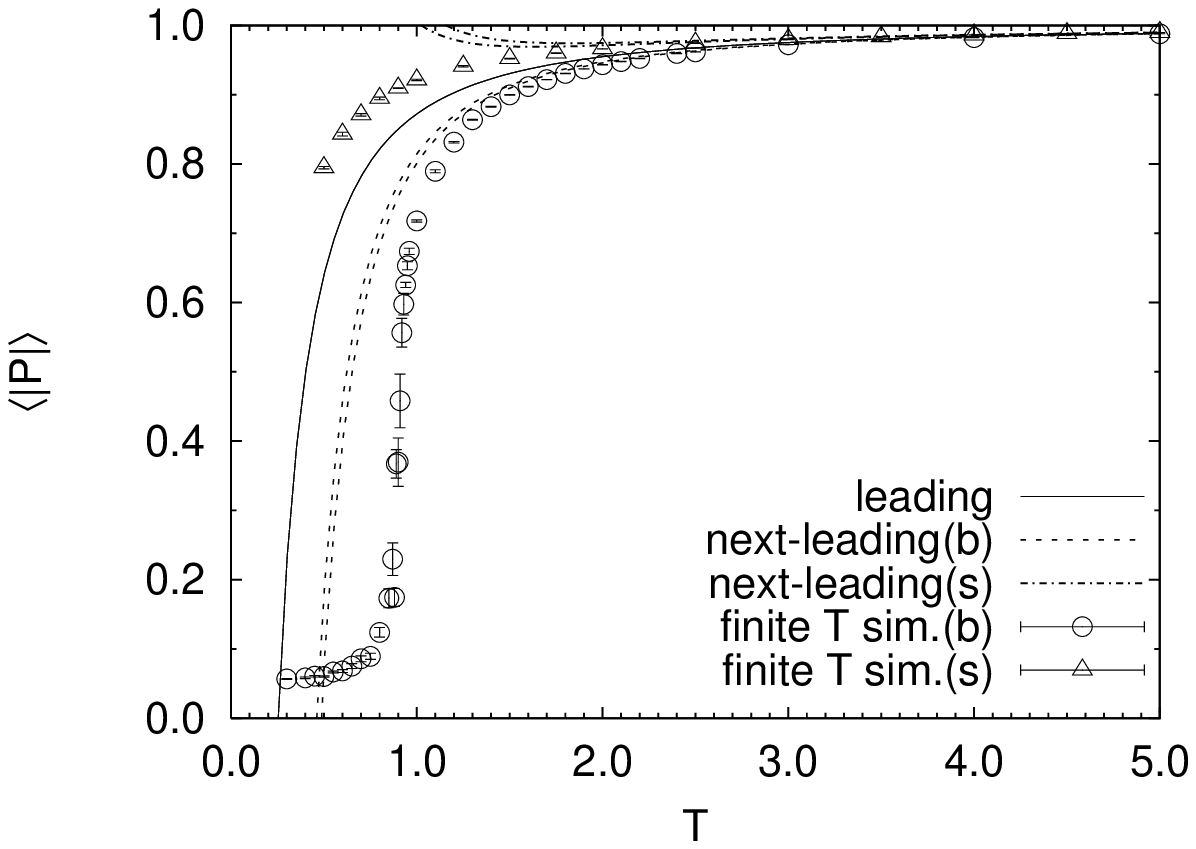,width=7.0cm}
\epsfig{file=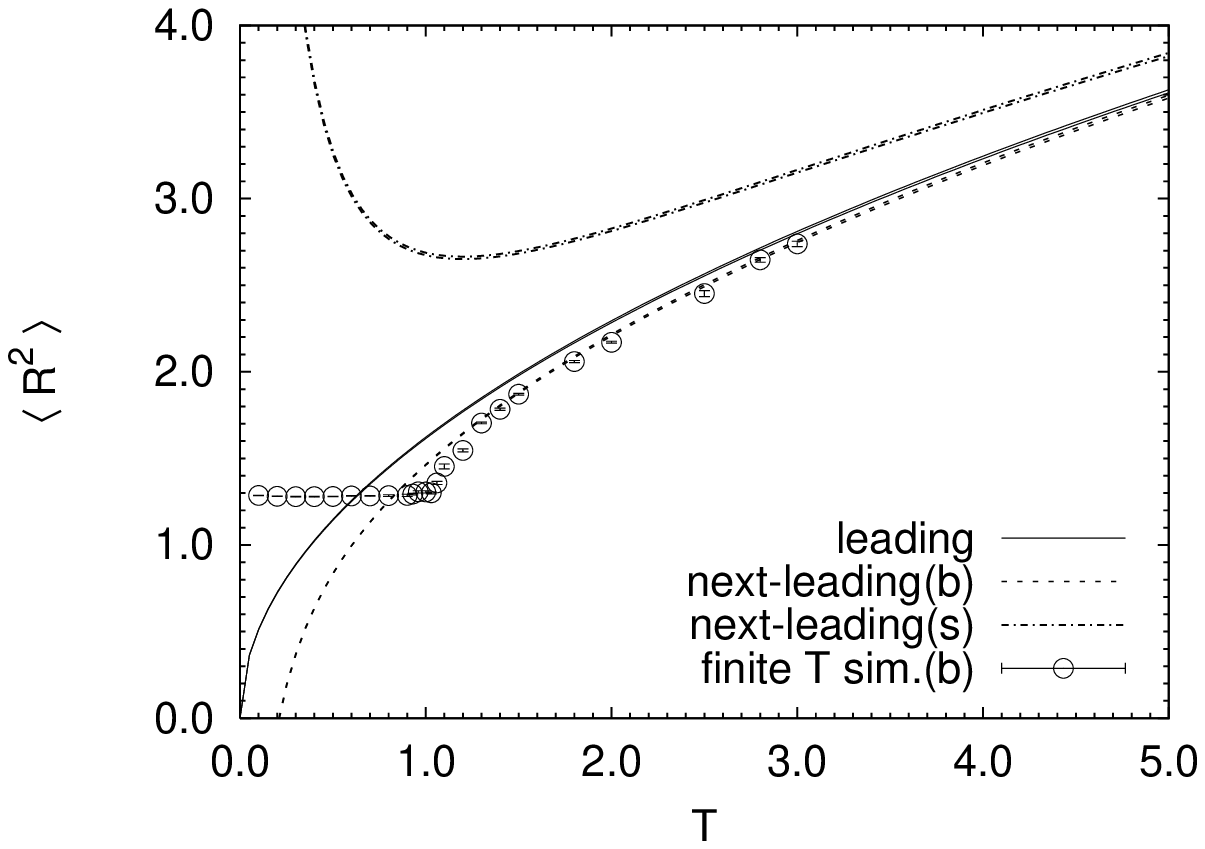,width=7.0cm} 
\epsfig{file=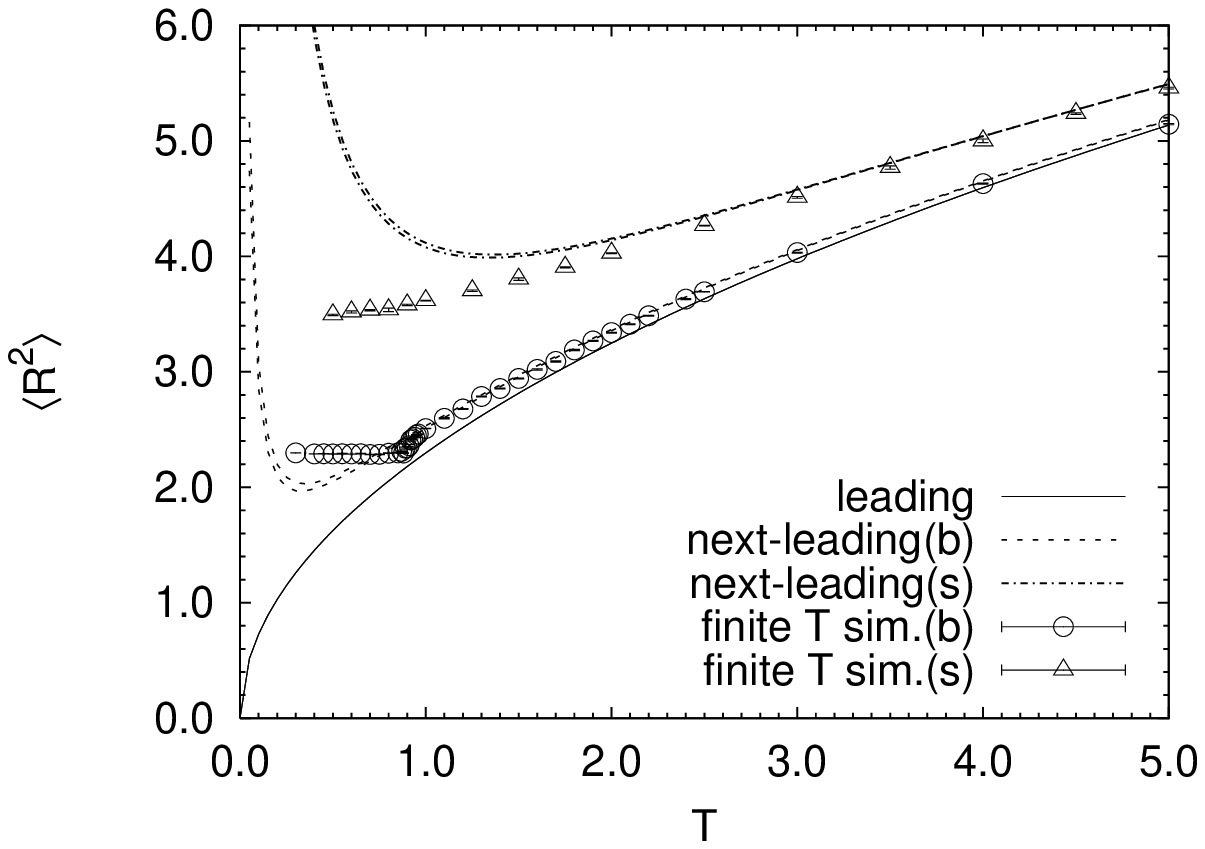,width=7.0cm} 
\caption{
Various quantities
are shown for
$d=3$ (left column) and $d=9$ (right column).
The curves represent the results obtained
by the high temperature expansion
using
the large-$N$ extrapolated values for $\chi_i$
shown in table \ref{Tab_chi5678}.  
The solid lines represent
the leading order results, which 
are the same for 
%do not distinguish
%(They are the same
%to 
%for 
the bosonic and supersymmetric cases.
The dashed lines and the dash-dotted lines
represent the next-leading order results
for the bosonic case
and the supersymmetric case, respectively.
The circles and squares
represent the results 
obtained by Monte Carlo simulation at finite $T$
for the bosonic model with $N=16$ 
\cite{Kawahara:2007nw,Kawahara:2007fn}
and for the supersymmetric model with 
$N=12$ \cite{Anagnostopoulos:2007fw}, respectively.
%%The plots on the left and right 
%%columns shows the results for 
%%$N=4$ and 
%The solid lines represent the leading order results,
%and the dashed lines and the dash-dotted lines represent
%the results including the next-leading order terms
%for the bosonic case and the supersymmetric case, respectively. 
%The circles represent the results for the bosonic case
%obtained by Monte Carlo simulation of the full model. 
}     
\label{Fig_HTE d3}} 

%==================================================================
\section{Summary and discussions}
\label{sec:Summary}

In this paper we have formulated the high temperature
expansion for the supersymmetric matrix quantum mechanics
with 4, 8 and 16 supercharges.
While the non-zero modes become weakly coupled
at high temperature,
the zero modes remain strongly coupled
and hence they have to be treated non-perturbatively.
This makes the problem nontrivial, but
we are able to obtain the next-leading
order terms by evaluating connected Green's function
in the bosonic IKKT model using Monte Carlo simulation.
Since the fermions decouple at the leading order, 
it is highly motivated to carry out the next-leading order
calculation.
Indeed, our results including the next-leading
order terms are in good agreement with the finite temperature
calculations down to $T\simeq 2$ in units of the
't Hooft coupling constant.
Note also that Monte Carlo evaluation
of the connected Green's functions in the bosonic IKKT model 
is by far easier than simulating the supersymmetric
matrix quantum mechanics at finite temperature directly.
This enables us to study the behavior at larger $N$
and to make a reliable large-$N$ extrapolation.
Our results confirm that the values of $N$
used in finite temperature simulations
are already large
enough to probe the 't Hooft large-$N$ limit 
at high temperature.

It is straightforward to extend our calculation to
higher orders.
% in the high temperature expansion. 
For that, one needs to
evaluate connected Green's functions
with more than two operators inserted.
That will require more statistics in
Monte Carlo evaluation. For finite $N$,
it is anticipated that the connected Green's functions
with many insertions of the $\tr (\tilde{A}_\mu)^2$ operator
would be divergent \cite{Krauth:1999qw},
and the order at which such divergence shows up would
grow linearly with $N$.
This property of the high temperature expansion
is reminiscent of the infrared instability observed in
Monte Carlo simulation of the supersymmetric model
at finite temperature \cite{Anagnostopoulos:2007fw}.
We note, however, 
that the divergence in the 
high temperature expansion occurs also in the bosonic case,
in which the finite temperature
%$T$ 
calculations exhibit 
no such instability.

%% We first integrate out the non-zero modes
%% perturbatively, and the remaining integration over
%% the zero modes have been done by Monte Carlo simulation.
%% Since fermions do not contain zero modes, they are
%% are integrated perturbatively. The Monte Carlo simulation

%% , which reveal substantial effects of fermions,
%% which 
%% We have performed large-$N$ extrapolation 
%% using explicit results for the coefficients
%% at $N=16,20,32$.

%Given the difficulty in increasing $N$ in the
%finite-$T$ simulations including fermions,
%we consider that our results provide useful
%information concerning the large-$N$ limit.
%The agreement with the finite-$T$ simulations
%at $N=12$, in turn 

It is worth while
%would be interesting 
to generalize our formulation
to higher dimensions.
For instance,
an interesting phase structure is expected
in 2d U($N$) $\mathcal{N}=8$
super Yang-Mills theory on a finite 
torus.\footnote{The situation becomes more involved
in dimensions higher than two.
Recently cascade transitions
from the black $p$-brane solution to 
the black $(p-1)$-brane solution have been found
in the dual gravity theories \cite{Hanada:2007wn}.
On the other hand, the high temperature limit of the 
4d U($N$) $\mathcal{N}=4$
super Yang-Mills theory on a finite torus,
for instance, is described by the
dimensionally reduced
3d bosonic model.
Analogous cascade transitions 
were observed earlier in the large-$N$ 
pure Yang-Mills theory on a 3d torus \cite{NN}
and in a related model \cite{Hanada:2006vm}.}
%Certainly there are lots of things to be clarified.}
%which can be obtained by dimensionally reducing
%dimensionally reducing
%${\cal N}=1$ U($N$) super Yang-Mills theory in $D=10$ 
%to 2 dimension,
In the strong coupling and low temperature regime,
the gauge/gravity duality predicts \cite{Aharony:2004ig,HO}
that there exists a phase transition corresponding to
the black-string/black-hole transition \cite{Gregory:1993vy}
in the dual gravity theory.
In the weak coupling and high temperature regime,
on the other hand,
one can study the theory by dimensionally reduced
1d bosonic model.
In ref.\ \cite{Aharony:2004ig},
the phase transition observed in the latter regime
has been 
conjectured to be a continuation of the 
phase transition in the former regime.
%mentioned above.
%black-string/black-hole transition in the dual gravity theory.
In ref.\ \cite{Kawahara:2007fn} the critical region
of the dimensionally reduced 1d bosonic model
has been studied more carefully, and a new phase
characterized by the non-uniform eigenvalue distribution
of the holonomy matrix has been discovered.
%So far, the calculations in the latter regime includes
%only the leading terms in the high temperature expansion.
Of particular interest from the viewpoint of 
the gauge/gravity duality
is 
%it would be interesting 
to investigate the fate
of this new phase as one lowers the temperature.
Calculations including the next-leading order terms
in the high temperature expansion would be useful for
such purposes.
%By including the effects of the next-leading order terms,
%one would be able to discuss 

%\vspace*{0.5cm}
\section*{Acknowledgements} 

We would like to thank
Masanori Hanada for valuable comments and 
discussions.
The simulations were performed on the PC clusters 
%with 25 nodes of Pentium4 (2.)
at KEK.

\appendix

\section*{Appendix}

%==================================================================
\section{Derivation of the formula for the internal energy}
\label{sec:EO}
%==================================================================

In this Appendix we derive the formula (\ref{def E}),
which is used to calculate the internal energy 
by the high temperature expansion.
%The derivation in 
The case without fermions is given 
in ref.\ \cite{Kawahara:2007fn}.
Let us first rewrite (\ref{defE}) as
\beq
E = - \frac{1}{Z(\beta)}
\lim_{\Delta \beta \rightarrow 0 } 
\frac{Z(\beta ') - Z(\beta)}{\Delta \beta} \ ,
\label{E-cal}
\eeq
where $\beta ' = \beta + \Delta \beta$,
and represent $Z(\beta ')$ for later convenience as
\beq
Z(\beta ') = \int 
[{\cal D} A']_{\beta '}
[{\cal D} X ']_{\beta '}
[{\cal D} \psi ']_{\beta '}
 \, \ee^{- S'}  \ ,
\eeq
where $S'$ is obtained from $S$ given in (\ref{action})
by replacing $\beta$, $t$, $A(t)$, $X_i(t)$, $\psi_\alpha(t)$
with $\beta '$, $t'$, $A'(t')$, $X_i '(t')$, $\psi_\alpha ' (t')$.
In order to relate $Z(\beta ')$ to $Z(\beta)$,
we consider the transformation
\beq
t ' = \frac{\beta '}{\beta} \, t \ ,
\quad
A '(t ') = \frac{\beta}{\beta '} \, A(t) \ ,
\quad
X_i ' (t ') = \sqrt{\frac{\beta '}{\beta}} \, X_i (t) \ ,
\quad
\ps'(t') = \ps(t) \ ,
\eeq
where the constant factors are
motivated on dimensional grounds, and we have
%in particular
$ [ {\cal D} X ' ]_{\beta ' } = [ {\cal D} X]_{\beta}$,
$ [ {\cal D} \psi ' ]_{\beta ' } = [ {\cal D} \psi]_{\beta}$
and
$[ {\cal D} A ' ]_{\beta ' } = [ {\cal D} A]_{\beta} $.
Under this transformation, 
the kinetic term in $S'$ reduces to that in $S$,
but the interaction term transforms non-trivially as
%% Note also that the kinetic term in the action
%% (\ref{action}) is invariant under this transformation.
%% The interaction term is not invariant, though,
%% and we obtain
\beqa
%--------------------------------------------------
\int_0^{\beta '} \!\!dt ' \, \tr
\Bigl( [X_i ' (t ' ),X_j ' (t ' )] \Bigr)^2 
&=& 
\left(\frac{\beta'}{\beta}\right)^3
\int_0^\beta  \!\!dt  \, \tr
\Bigl( [X_i  (t  ),X_j (t )] \Bigr)^2 \ ,\\
%--------------------------------------------------
\int_0^{\beta '}  \!\!dt  \, 
\tr 
\Bigl( \ps_\alpha (t ' ) 
% (\gm_i)_{\alpha\beta} 
[{X_i}' (t ' ) ,\ps'_\beta (t ' )] \Bigr)
&=& 
\left(\frac{\beta'}{\beta}\right)^{3/2}
\int_0^\beta  \!\!dt  \, 
\tr
\Bigl( \ps_\alpha 
%(\gm_i)_{\alpha\beta} 
[X_i(t),\ps_\beta(t)] \Bigr) \ .
%--------------------------------------------------
\eeqa
This gives us the relation
\beqa
\label{Z/Z}
Z(\beta')=Z(\beta) \left(
1 -  N^2 \Delta \beta  
\left(
{\cal E}_{\rm b} + {\cal E}_{\rm f} 
\right)
+{\rm O}((\Delta \beta)^2)
\right) \ ,
\eeqa 
where the coefficients ${\cal E}_{\rm b}$ 
and ${\cal E}_{\rm f}$ are defined by (\ref{def-Eb})
and (\ref{def-Ef}).
Plugging these into (\ref{E-cal}), we get (\ref{def E}).
Thus we are able to express the internal energy $E$
in terms of the expectation values,
which can be calculated directly by Monte Carlo simulation. 

%==================================================================
\section{Increasing statistics by exploiting SO($D$) symmetry}
\label{sec:Repre} 

In eqs.\ (\ref{DR1}) and (\ref{DR2}),
we have rewritten the expectation values
that define $\chi_1$ and $\chi_2$
by exploiting the SO($D$) symmetry of the DR model (\ref{DRmodel}).
Similar rewriting can be done also
for the other coefficients $\chi_i \ (i=3,\cdots,8)$
defined in 
%(\ref{DR1}), 
eqs.\ (\ref{chi3-4}),
(\ref{chi5-6}) and (\ref{chi7-8})
as presented below.
%and we give the explicit form in the following.
In actual measurements in the Monte Carlo simulation,
we can increase the statistics considerably
by using these forms instead of the original ones.
\beqa
%\chi_1&=&\frac{d}{D}\Big\langle \frac{1}{N} 
%{\rm tr}\tilde{A}_\mu{}^2 \Big\rangle_{\rm DR} \ , \nonumber \\ 
\chi_3&=&     
\frac{d}{D} \sum_\mu \Big\langle \tr
(\tilde{A}_\mu)^2 \cdot \tr
(\tilde{A}_\mu)^2 \Big\rangle_{\rm DR,C} 
 +  \frac{2{}_d {\rm C}_2}{ {}_D {\rm C}_2} 
\sum_{\mu<\nu} \Big\langle \tr (\tilde{A}_\mu)^2 
\cdot \tr (\tilde{A}_\nu)^2 
\Big\rangle_{\rm DR,C} \ , \nonumber \\ 
\chi_4&=& \frac{d}{{}_D {\rm C}_2} 
\sum_{\mu<\nu} \Big\langle \tr 
(\tilde{A}_\mu)^2 \cdot \tr(\tilde{A}_\nu)^2 
\Big\rangle_{\rm DR,C} \ , \nonumber \\ 
\chi_5&=&  \frac{2{}_d {\rm C}_2}{{}_D {\rm C}_2} 
\sum_{\mu<\nu}
\Big\langle \tr (\tilde{F}_{\mu\nu})^2 
\cdot \tr
\Big\{ (\tilde{A}_\mu)^2 + (\tilde{A}_\nu)^2 \Big\}
 \Big\rangle_{\rm DR,C} \nonumber \\
&~& +  \frac{2(d-2){}_d {\rm C}_2}{(D-2){}_D {\rm C}_2} 
\sum_{\mu<\nu} \sum_{\lambda \not= \mu,\nu} 
\Big\langle \tr (\tilde{F}_{\mu\nu})^2 
\cdot \tr (\tilde{A}_\lambda)^2 
\Big\rangle_{\rm DR,C} \ , 
\nonumber\\ 
\chi_6&=&\frac{2{}_d {\rm C}_2}{(D-2){}_D {\rm C}_2} 
\sum_{\mu<\nu} \sum_{\lambda \not= \mu,\nu} 
\Big\langle \tr (\tilde{F}_{\mu\nu})^2 
\cdot \tr (\tilde{A}_\lambda)^2 
\Big\rangle_{\rm DR,C} \ , 
\nonumber \\ 
\chi_7&=&\frac{1}{D} \sum_\mu 
\Big\langle \tr (\tilde{A}_\mu)^2 \cdot 
\tr (\tilde{A}_\mu)^2 \Big\rangle_{\rm DR,C} \ , 
\nonumber \\ 
\chi_8&=&\frac{1}{D} \Big\langle \frac{1}{N} 
\tr (\tilde{A}_\mu)^4 \Big\rangle_{\rm DR} \ .
\eeqa

%==================================================================

\begin{table}
\begin{center}
\begin{tabular}{|c|c|c|c|c|c|c|c|c|}
\hline 
$d$ & $N$ & $\chi_1$ & $\chi_3$ & $\chi_4$ & $\chi_5$ & $\chi_6$ & $\chi_7$ & $\chi_8$  \\
\hline \hline  
3 &  4       & 1.48(1)    & 2.79(8)  & -0.48(5)   & 3.29(1)  & -0.70(4)   & 1.40(5)   & 0.60(1)    \\ \hline
3 &  8       & 1.6579(4)  & 2.756(7) & -0.751(4)  & 3.294(6) & -1.162(5)  & 1.419(4)  & 0.6581(4)  \\ \hline
3 & 10       & 1.6305(3)  & 2.56(1)  & -0.718(9)  & 3.32(1)  & -1.10(1)   & 1.33(1)   & 0.6342(9)  \\ \hline
3 & 12       & 1.6295(3)  & 2.319(6) & -0.708(4)  & 3.195(7) & -1.140(6)  & 1.245(4)  & 0.6239(3)  \\ \hline
3 & 16       & 1.6229(1)  & 2.230(4) & -0.720(3)  & 3.148(5) & -1.102(5)  & 1.223(3)  & 0.6156(1)  \\ \hline
3 & 20       & 1.6198(1)  & 2.070(4) & -0.677(9)  & 2.970(1) & -1.04(1)   & 1.142(7)  & 0.6123(1)  \\ \hline
3 & 32       & 1.61697(4) & 1.940(6) & -0.633(3)  & 2.797(9) & -0.981(5)  & 1.069(4)  & 0.60780(5) \\ \hline
3 & $\infty$ & 1.6150(1)  & 1.83(1)  & -0.6039(6) & 2.676(9) & -0.940(2)  & 1.016(5)  & 0.6051(2)  \\
\hline \hline
5 &  4       & 1.821(1)   & 1.56(1)  & -0.209(3)  & 3.53(1)  & -0.732(8) & 0.481(4)  & 0.2812(8)  \\ \hline
5 &  8       & 1.8331(2)  & 1.181(2) & -0.2157(7) & 3.564(7) & -0.703(2) & 0.4087(8) & 0.2778(1)  \\ \hline
5 & 10       & 1.8356(6)  & 1.179(6) & -0.232(2)  & 3.68(1)  & -0.732(8) & 0.421(2)  & 0.2785(2)  \\ \hline
5 & 12       & 1.8377(3)  & 1.153(3) & -0.229(1)  & 3.633(9) & -0.703(2) & 0.414(1)  & 0.2788(1)  \\ \hline
5 & 16       & 1.83935(9) & 1.141(2) & -0.2251(9) & 3.658(7) & -0.723(2) & 0.4084(8) & 0.27893(3) \\ \hline
5 & 20       & 1.8387(1)  & 1.104(2) & -0.2220(8) & 3.544(6) & -0.718(2) & 0.3985(8) & 0.27866(4) \\ \hline
5 & 32       & 1.8393(3)  & 1.041(3) & -0.2282(5) & 3.35(1)  & -0.744(1) & 0.3909(7) & 0.27874(1) \\ \hline
5 & $\infty$ & 1.8382(8)  & 1.01(1)  & -0.229(3)  & 3.27(4)  & -0.751(9) & 0.384(1)  & 0.27868(5) \\
\hline \hline  
9 &  4       & 2.191(1)   & 0.769(5) & -0.0925(5) & 3.99(1) & -0.558(2) & 0.1681(6) & 0.1199(1)  \\ \hline
9 &  8       & 2.2700(2)  & 0.746(1) & -0.0861(3) & 4.34(1) & -0.510(1) & 0.1595(2) & 0.12894(3) \\ \hline 
9 & 10       & 2.2810(5)  & 0.766(3) & -0.0859(6) & 4.44(2) & -0.506(3) & 0.1615(5) & 0.13045(7)  \\ \hline
9 & 12       & 2.2854(3)  & 0.751(4) & -0.0863(6) & 4.44(2) & -0.510(1) & 0.1602(2) & 0.13114(3) \\ \hline 
9 & 16       & 2.2901(1)  & 0.746(2) & -0.0886(5) & 4.43(1) & -0.525(3) & 0.1617(2) & 0.13163(2) \\ \hline
9 & 20       & 2.2932(3)  & 0.734(3) & -0.0912(9) & 4.40(2) & -0.55(1)  & 0.1631(6) & 0.13204(3) \\ \hline
9 & 32       & 2.29566(7) & 0.730(6) & -0.082(1)  & 4.38(2) & -0.59(1)  & 0.1399(1) & 0.13234(1) \\ \hline  
9 & $\infty$ & 2.2975(1)  & 0.719(6) & -0.082(6)  & 4.36(2) & -0.61(2)  & 0.14(2)   & 0.13257(5) \\
\hline
\end{tabular} 
\end{center} 
\caption{
The values of $\chi_i \ (i=1,3,\cdots,8)$ for various $d$
and $N$ obtained by Monte Carlo simulation of the
corresponding DR model.
The values at $N=\infty$ are obtained
by extrapolating the results for $N=16,20,32$ 
as shown in fig.\ \ref{Fig_HTE data}.
% 
% and $N$
%obtained by Monte Carlo simulation of the dimensionally
%reduced model.
} 
\label{Tab_chi5678}  
\end{table}

\end{document}